\documentclass[submission, Phys]{SciPost}

\usepackage[utf8]{inputenc} 
\usepackage[T1]{fontenc} 	
\usepackage[english]{babel} 

\usepackage[bitstream-charter]{mathdesign}
\usepackage{geometry} 		
\usepackage{microtype} 	
\usepackage{amsmath} 		
\usepackage{mathtools} 		
\usepackage{float} 			
\usepackage{graphicx} 		
\usepackage{tabularx} 		
\usepackage{booktabs} 		
\usepackage{color, xcolor} 	
\definecolor{Rcolor}{HTML}{E99595}
\definecolor{Gcolor}{HTML}{C5E0B4}
\definecolor{Bcolor}{HTML}{9DC3E6}
\definecolor{Ycolor}{HTML}{FFE699}
\definecolor{Vcolor}{HTML}{CFB4E0}
\usepackage{pdfpages} 		
\usepackage{extarrows} 		
\usepackage{multirow} 		
\usepackage{multicol} 		
\usepackage{enumitem} 		
\usepackage{xspace} 		
\usepackage{stackrel} 		
\usepackage{tikz} 			
\usetikzlibrary{shapes.arrows,arrows.meta,fit}
\usetikzlibrary{positioning,calc}
\usepackage{braket} 		
\usepackage{bm} 			
\usepackage{tensor} 		
\usepackage{slashed} 		
\usepackage{siunitx} 		
\usepackage{lastpage} 		
\usepackage{cite} 			
\usepackage[normalem]{ulem} 
\usepackage{fontawesome} 	
\usepackage{tocloft} 		
\usepackage{titlesec} 		
\usepackage{doi} 			
\usepackage{hyperref} 		
\usepackage[most]{tcolorbox} 					
\usepackage[nameinlink, capitalize]{cleveref} 	
\crefname{section}{Sec.}{Secs.}
\crefname{subsection}{Sec.}{Secs.}
\crefname{subsubsection}{Sec.}{Secs.}
\crefname{appendix}{App.}{Apps.}
\crefname{table}{Tab.}{Tabs.}
\Crefname{section}{Section}{Sections}
\Crefname{subsection}{Section}{Sections}
\Crefname{subsubsection}{Section}{Sections}
\Crefname{appendix}{Appendix}{Appendices}
\Crefname{table}{Table}{Tables}
\crefformat{equation}{#2Eq.(#1)#3}
\crefmultiformat{equation}{#2Eqs.(#1)#3}{ and #2(#1)#3}{, #2(#1)#3}{ and #2(#1)#3}
\crefrangeformat{equation}{#3Eqs.(#1)#4--#5(#2)#6}
\Crefformat{equation}{#2Equation~(#1)#3}
\Crefmultiformat{equation}{#2Equations~(#1)#3}{ and #2(#1)#3}{, #2(#1)#3}{ and #2(#1)#3}
\Crefrangeformat{equation}{#3Equations~(#1)#4--#5(#2)#6}

\usepackage[nottoc, notlot, notlof]{tocbibind} 	
\usepackage[ruled, vlined]{algorithm2e} 		
\usepackage{makecell}
\usepackage[makeroom]{cancel}
\usepackage{feynmf}
\usepackage[compat=1.1.0]{tikz-feynman}
\ifluatex\else
  \tikzfeynmanset{
    warn luatex=false,
    every medium@@/.style={/tikz/node distance=1.5cm}
  }
\fi

\makeatletter
\def\BState{\State\hskip-\ALG@thistlm}
\makeatother

\makeatletter
\@ifundefined{pdfoutput}{}{\DeclareGraphicsRule{*}{mps}{*}{}}
\makeatother

\makeatletter
\DeclareRobustCommand*{\bfseries}{%
   \not@math@alphabet\bfseries\mathbf
   \fontseries\bfdefault\selectfont
   \boldmath
}
\makeatother

\hypersetup{
	pdftitle={Production-ready MadNIS at LO},
	pdfauthor={Heimel et al.},
	colorlinks=true, 			
	linkcolor={red!50!black}, 	
	citecolor={blue!50!black}, 	
	urlcolor={blue!80!black} 	
} 

\DeclareSymbolFont{usualmathcal}{OMS}{cmsy}{m}{n}
\DeclareSymbolFontAlphabet{\mathcal}{usualmathcal}

\SetArgSty{textnormal}
\SetKwComment{Comment}{{\small\#}~}{}
\SetCommentSty{mycommfont}

\setitemize{itemsep=0pt, parsep=0pt} 				
\setenumerate{itemsep=0pt, parsep=0pt} 				
\setitemize{itemsep=2pt,topsep=2pt,parsep=0pt,partopsep=0pt,leftmargin=*}
\setenumerate{itemsep=0pt,topsep=2pt,parsep=0pt,partopsep=0pt,labelindent=3pt,leftmargin=*}
\newlist{todolist}{itemize}{2}
\setlist[todolist]{label=$\square$}
\usepackage{pifont}

\usepackage{amsmath}
 
\usepackage{amsthm} 		
\theoremstyle{definition}

\definecolor{red_cb}{HTML}{e41a1c}
\definecolor{blue_cb}{HTML}{377eb8}
\definecolor{green_cb}{HTML}{4daf4a}
\definecolor{purple_cb}{HTML}{984ea3}
\definecolor{orange_cb}{HTML}{ff7f00}

\definecolor{EmeraldGreen}{HTML}{1ea78d}
\definecolor{EnglishRed}{HTML}{b02427}
\hypersetup{colorlinks=true,urlcolor=EmeraldGreen,citecolor=EmeraldGreen,linkcolor=EnglishRed}

\newcommand{\ie}{\text{i.e.}\;}

\newcommand{\gbar}{\overline{G}}

\newcommand{\mwith}{\text{with}}
\newcommand{\mand}{\text{and}}

\newcommand{\qqquad}{\qquad\quad}

\def\d{\mathrm{d}}

\newcommand\one{\leavevmode\hbox{\small1\normalsize\kern-.33em1}}
\newcommand{\imag}{\mathrm{i}} 				

\newcommand{\loss}{\mathcal{L}} 	

\newcommand{\vegas}{\texttt{VEGAS}\xspace}

\newcommand{\mg}{\texttt{MadGraph}\xspace}
\newcommand{\mgfive}{\texttt{MG5aMC}\xspace}

\newcommand{\mgseven}{\texttt{MadGraph7}\xspace}
\newcommand{\pythia}{\texttt{PYTHIA}\xspace}
\newcommand{\madspace}{\texttt{MadSpace}\xspace}

\newcommand{\sherpa}{\texttt{Sherpa}\xspace}
\newcommand{\pepper}{\texttt{Pepper}\xspace}
\newcommand{\herwig}{\texttt{Herwig}\xspace}

\newcommand{\madnis}{\texttt{MadNIS}\xspace}
\newcommand{\madmatrix}{\texttt{MadMatrix}\xspace}
\newcommand{\amplicol}{\texttt{AmpliCol}\xspace}
\newcommand{\whizard}{\texttt{WHIZARD}\xspace}
\newcommand{\lhapdf}{\texttt{LHAPDF6}\xspace}

\newcommand{\ttrain}{t_\text{train}}

\newcommand{\arXiv}[2][]{%
	\ifthenelse{\equal{#1}{}}%
	{\href{http://arxiv.org/abs/#2}{arXiv:#2}}%
	{\href{http://arxiv.org/abs/#2}{arXiv:#2~[#1]}}}

\def\slashchar#1{\setbox0=\hbox{$#1$}           
   \dimen0=\wd0                                 
   \setbox1=\hbox{/} \dimen1=\wd1               
   \ifdim\dimen0>\dimen1                        
      \rlap{\hbox to \dimen0{\hfil/\hfil}}      
      #1                                        
   \else                                        
      \rlap{\hbox to \dimen1{\hfil$#1$\hfil}}   
      /                                         
   \fi}

\newcommand{\tikznode}[2]{%
\ifmmode%
\tikz[remember picture,baseline=(#1.base),inner sep=0pt] \node (#1) {$#2$};%
\else
\tikz[remember picture,baseline=(#1.base),inner sep=0pt] \node (#1) {#2};%
\fi}

\def\mathswitchr#1{\relax\ifmmode{\mathrm{#1}}\else$\mathrm{#1}$\xspace\fi}
\def\mathswitch#1{\relax\ifmmode#1\else$#1$\xspace\fi}

\newcommand{\Pq}{\mathswitch q}
\newcommand{\Pqbar}{\mathswitch{\bar q}}
\newcommand{\PW}{\mathswitchr W}
\newcommand{\PWp}{\mathswitchr {W^+}}
\newcommand{\PWm}{\mathswitchr {W^-}}
\newcommand{\PWpm}{\mathswitchr {W^\pm}}
\newcommand{\PZ}{\mathswitchr Z}

\newcommand{\Pg}{\mathswitchr g}
\newcommand{\PH}{\mathswitchr H}

\newcommand{\Pep}{\mathswitchr {e^+}}
\newcommand{\Pem}{\mathswitchr {e^-}}

\newcommand{\Pp}{\mathswitchr p}
\newcommand{\Pd}{\mathswitchr d}
\newcommand{\Pu}{\mathswitchr u}

\newcommand{\Pb}{\mathswitchr b}
\newcommand{\Pt}{\mathswitchr t}
\newcommand{\Pj}{\mathswitchr j}
\newcommand{\Ptbar}{\mathswitchr{\bar t}}
\newcommand{\Pubar}{\mathswitchr{\bar u}}
\newcommand{\Pdbar}{\mathswitchr{\bar d}}

\newcommand{\Pl}{\ell}

\newcommand{\MZ}{\mathswitch {M_\PZ}}

\graphicspath{{./figs/}}

\begin{document}

\begin{center}{
\Large \textbf{\color{scipostdeepblue}{
MadNIS, Ready for Production at LO
}}}
\end{center}

\begin{center}
Theo Heimel\textsuperscript{1},
Olivier Mattelaer\textsuperscript{1},
Daniele Massaro\textsuperscript{2},
and Ramon Winterhalder\textsuperscript{3}
\end{center}

\begin{center}
{\bf 1} CP3, Universit\'e catholique de Louvain, Louvain-la-Neuve, Belgium
\\
{\bf 2} CERN, Geneva, Switzerland
\\
{\bf 3} TIFLab, Universit\`a degli Studi di Milano \& INFN Sezione di Milano, Italy
\end{center}

\begin{center}
\today
\end{center}


\section*{\color{scipostdeepblue}{Abstract}}
{\bf
We extend neural multi-channel importance sampling in MadNIS to full proton-proton processes in a production-ready event-generation setup.
We benchmark a wide range of LHC processes, including multi-jet, vector-boson, top-quark, and Higgs production, and obtain event-generation speed-ups of up to a factor of 250 without process-specific hyperparameter tuning.
We quantify how the speed-up scales with training time and final-state multiplicity, and when the training cost amortizes.
MadNIS will become the default for phase-space integration and event generation at leading order in MadGraph.
}

\vspace{10pt}
\noindent\rule{\textwidth}{1pt}
\tableofcontents\thispagestyle{fancy}
\noindent\rule{\textwidth}{1pt}
\vspace{10pt}

\clearpage
\section{Introduction}
\label{sec:intro}

Event generators such as \pythia~\cite{Bierlich:2022pfr}, \sherpa~\cite{Sherpa:2024mfk}, \herwig~\cite{Bellm:2025pcw}, and \mg~\cite{Maltoni:2002qb,Alwall:2007st,Alwall:2011uj,Alwall:2014hca,Frederix:2018nkq} turn a Lagrangian into simulated scattering events and provide the backbone of the LHC analysis chain~\cite{Campbell:2022qmc,vanBeekveld:2026uxl}. 
Their first step, the integration over parton-level phase space and the generation of unweighted events, already dominates the cost for high final-state multiplicities~\cite{Hoche:2019flt,CERN-LHCC-2022-005}. 
For the HL-LHC, the required samples and process complexity will grow faster than the available computing budget, so this step risks becoming the bottleneck of the entire simulation pipeline.

Modern machine learning~\cite{Butter:2022rso,Plehn:2022ftl} offers a way out, and by now accelerates essentially every step of the simulation chain~\cite{Feickert:2021ajf,HEP-ML-Zenodo, Krause:2026ayh}.
For the evaluation of scattering amplitudes, neural surrogates~\cite{Bishara:2019iwh,Badger:2020uow,Aylett-Bullock:2021hmo,Maitre:2021uaa,Danziger:2021eeg,Winterhalder:2021ngy,Janssen:2023ahv,Maitre:2023dqz,Brehmer:2024yqw,Breso-Pla:2024pda,Herrmann:2025nnz,Favaro:2025pgz,Villadamigo:2025our,Bahl:2026jvt} replace the expensive exact evaluation, including a calibrated uncertainty estimate~\cite{Badger:2022hwf,Bahl:2024gyt,Bahl:2025xvx,Bahl:2026qaf,Beccatini:2025tpk}.
For phase-space integration, neural importance
sampling~\cite{Bendavid:2017zhk,Klimek:2018mza,Chen:2020nfb,Gao:2020vdv,Bothmann:2020ywa,Gao:2020zvv,Deutschmann:2024lml,Janssen:2025zke,Bothmann:2025lwg} replaces or refines the classic adaptive mappings of
\vegas~\cite{Lepage:1977sw,Lepage:1980dq,Lepage:2020tgj} with normalizing flows. 
For \mg, this is realized by the \madnis framework~\cite{Heimel:2022wyj,Heimel:2023ngj,Heimel:2024wph}, which refines the channel weights of the single-diagram-enhanced multi-channel
decomposition~\cite{Maltoni:2002qb,Mattelaer:2021xdr}, and improves the phase-space sampling, while a first NLO version combines neural importance sampling with fast surrogates~\cite{DeCrescenzo:2026tsp}.
Neural control variates provide a complementary handle on the variance, reducing the spread of event weights and avoiding negative weights at LO and NLO~\cite{Heimel:2026cxh}.

Hardware acceleration is the second lever. GPU-based amplitude evaluation~\cite{Bothmann:2021nch} has matured into full parton-level event generation with \pepper~\cite{Bothmann:2023gew}.
Within \mg, the CUDACPP plugin, now dubbed \madmatrix, provides substantial speed-ups for the matrix-element evaluation~\cite{Valassi:2023yud,Hageboeck:2023blb,Wettersten:2025hrb,Valassi:2025xfn,Hagebock:2025jyk}, and \madspace~\cite{Heimel:2026hgp} provides the corresponding GPU-native phase-space generation.
Both levers combine naturally, because we train and evaluate the sampler on exactly the hardware that accelerates the matrix elements.
For \pepper, an end-to-end GPU workflow with normalizing-flow proposals recently achieved speed-ups of up to two orders of magnitude for high-multiplicity processes, including the training cost~\cite{Bothmann:2026yzp}.

So far, however, \madnis has delivered these gains only for single partonic subprocesses of selected benchmark processes.
In this paper, we extend \madnis to generic proton-proton processes, and make neural importance sampling the default in \mg at leading order. All developments are publicly available in the alpha release of \mgseven~\cite{MG7:v0.2.1}.
We generalize \madnis to include discrete degrees of freedom, allowing importance sampling of symmetry permutations and parton-flavor configurations.
We merge matrix-element implementations, integration channels of similar topology, and entire subprocesses, which reduces the number of independent trainings. 
We extend buffered training to unweighted samples, which stabilizes the training and allows faster convergence, while a decoupling of training and event generation further improves the throughput of the overall pipeline.
On top of these technical improvements, we define automated defaults, which set the hyperparameters from a short \vegas survey of each process.
Because every change of process, cuts, or model parameters requires a new training, the training time adds directly to the cost of event generation.
We therefore quantify how the gain scales with training time and process multiplicity, and when the training cost amortizes against the faster event generation.
Finally, we benchmark a broad range of LHC processes, from multi-jet and vector-boson production to top-quark and Higgs processes, vector-boson scattering, and decay chains, and we confirm that the automated setup still reproduces the correct reference cross sections and distributions.

This paper is organized as follows. In \cref{sec:recap} we briefly review multi-channel importance sampling and the \madnis method.
\Cref{sec:technical} introduces the improvements that make \madnis robust and scalable.
In \cref{sec:production} we study the training time and its amortization, validate the automated setup, and benchmark a broad range of LHC processes.
We end with a conclusion and outlook in \cref{sec:outlook}.

\clearpage
\section{Neural importance sampling with MadNIS}
\label{sec:recap}

To be self-contained, we briefly review multi-channel Monte Carlo integration and the \madnis method~\cite{Heimel:2022wyj,Heimel:2023ngj,Heimel:2024wph}.
Readers already familiar with both topics can proceed directly to \cref{sec:technical}. 
We start by considering the integral
of a function $f(\Phi_n)\sim\vert\mathcal{M}\vert^2$ over the $D$-dimensional $n$-body phase space,
\begin{align}
    I[f] = \int \d\Phi_n\,f(\Phi_n) \qquad \mwith \qquad 
    \Phi_n\subseteq\mathbb{R}^D \qquad D=3n-4 \; .
    \label{eq:psinteg}
\end{align}
Following the single-diagram-enhanced decomposition of \mg~\cite{Maltoni:2002qb,Mattelaer:2021xdr},
we split the integrand into $N_c$ channels with local channel weights $\alpha_i(\Phi_n)$,
\begin{align}
  f(\Phi_n)
  = \sum_{i=1}^{N_c} \alpha_i(\Phi_n)\,f(\Phi_n)
  \qquad \mwith \qquad \sum_{i=1}^{N_c} \alpha_i(\Phi_n) = 1
  \qquad \mand \qquad \alpha_i(\Phi_n) \ge 0 \; ,
  \label{eq:mc-decomp}
\end{align}
so each channel focuses on different integrand structures.
Similar decompositions~\cite{Kleiss:1994qy,Weinzierl:2000wd} relying on global, \ie non phase-space dependent, channel weights $\alpha_i$ are used within
\sherpa~\cite{Sherpa:2024mfk} and \whizard~\cite{Kilian:2007gr}. 
For details on the global- versus
local-weight variants and their algorithmic differences, we refer to Ref.~\cite{Heimel:2026hgp}.
Within each channel, we introduce an invertible mapping between the physical phase space and the
unit hypercube,
\begin{align}
\Phi_n \quad
\xleftrightarrow[\quad \leftarrow \gbar_i(x)\quad]{G_i(\Phi_n)\rightarrow}
\quad x \in [0,1]^D \; ,
\label{eq:ps-mapping}
\end{align}
which defines a normalized channel density
\begin{align}
   g_i(\Phi_n) = \left\vert\frac{\partial G_i(\Phi_n)}{\partial \Phi_n}\right\vert
   \qquad \mwith \qquad \int \d\Phi_n\,g_i(\Phi_n) = 1 \; .
   \label{eq:channel-density}
\end{align}
The mappings $G_i$ are constructed from the analytic structure of the amplitude, for instance
resonant propagators and soft or collinear limits, and implemented in \madspace~\cite{Heimel:2026hgp}.
Evaluating each channel with its own sampling density turns \cref{eq:psinteg} into the
multi-channel importance-sampling estimator
\begin{align}
  I[f]
  = \sum_{i=1}^{N_c} \int \d\Phi_n\;g_i(\Phi_n)\,\frac{\alpha_i(\Phi_n)\,f(\Phi_n)}{g_i(\Phi_n)}
  \equiv \sum_{i=1}^{N_c} \left\langle \frac{\alpha_i(\Phi_n)\,f(\Phi_n)}{g_i(\Phi_n)}
  \right\rangle_{\Phi_n\sim g_i(\Phi_n)} \; .
  \label{eq:mc-master}
\end{align}
Sampling from $g_i$ amounts to drawing $x$ uniformly and mapping onto phase space. 
It is therefore convenient to work directly on the unit hypercube and to absorb the analytic channel density into a reweighted channel integrand,
\begin{align}
  \bar f_i(x)
  = \frac{f(\Phi_n)}{g_i(\Phi_n)}\bigg\vert_{\Phi_n=\gbar_i(x)}\; ,
  \label{eq:fbar}
\end{align}
so that \cref{eq:mc-master} becomes a plain integral over the unit hypercube,
\begin{align}
  I[f] = \sum_{i=1}^{N_c} \int \d x\; \alpha_i\!\left(\gbar_i(x)\right)\,\bar f_i(x)
  \equiv \sum_{i=1}^{N_c}
  \Bigl\langle \alpha_i\!\left(\gbar_i(x)\right)\,\bar f_i(x) \Bigr\rangle_{x\sim \mathcal{U}}\; .
  \label{eq:mc-master-x}
\end{align}
Only the generation direction $\gbar_i$ enters, so the analytic mappings never have to be
inverted.

\madnis improves both ingredients of \cref{eq:mc-master-x} with neural networks. It learns a
correction to the channel weights $\alpha_i(\Phi_n)$, and it refines the analytic mappings $\gbar_i(x)$ by composing them with an additional learned transformation.

\subsubsection*{Neural channel weights}
 
The channel-weight network (CWnet) encodes the local weights, $\alpha_i(\Phi_n)\to\alpha_{i,\xi}(\Phi_n)$ with parameters $\xi$. 
The normalization and positivity constraint of \cref{eq:mc-decomp} is built into the architecture through a softmax, \ie exponential normalization, over the unnormalized network output $A_{i,\xi}(\Phi_n)$~\cite{Heimel:2022wyj},
\begin{align}
  \alpha_{i,\xi}(\Phi_n) = \frac{\exp A_{i,\xi}(\Phi_n)}{\sum_j \exp A_{j,\xi}(\Phi_n)} \; .
  \label{eq:alpha-net}
\end{align}
Because the weights vary strongly across phase space, we do not learn them from scratch but as a correction to a physics-motivated prior. 
Two well-motivated, normalized priors define the single-diagram-enhanced method of \mg~\cite{Maltoni:2002qb,Mattelaer:2021xdr},
\begin{align}
\begin{split}
    \text{SDE-1:}\qquad \alpha_i^{\text{MG}}(\Phi_n) &= \frac{\vert\mathcal{M}_i(\Phi_n)\vert^2}{\sum_j \vert\mathcal{M}_j(\Phi_n)\vert^2} \\
    \text{SDE-2:}\qquad \alpha_i^{\text{MG}}(\Phi_n) &= \frac{P_i(\Phi_n)}{\sum_j P_j(\Phi_n)}
    \qquad \mwith \qquad
    P_i(\Phi_n) = \prod_{k \in \text{prop}} \frac{1}{\vert p_k^2 - m_k^2 - \imag\,m_k\Gamma_k\vert^2} \; ,
\end{split}
\label{eq:mg-prior}
\end{align}
built either from single-diagram matrix elements or from products of propagator denominators.
Relative to either prior we learn only a correction, \ie
\begin{align}
  \alpha_{i,\xi}(\Phi_n)
  = \frac{\alpha^{\text{MG}}_i(\Phi_n)\,\exp A_{i,\xi}(\Phi_n)}
         {\sum_j\left[\alpha^{\text{MG}}_j(\Phi_n)\,\exp A_{j,\xi}(\Phi_n)\right]} \; ,
  \label{eq:alpha-correction}
\end{align}
and initialize the parameters such that $A_{i,\xi_0}(\Phi_n)=0$, so training starts exactly from the \mg prior. 
Note that $A_{j,\xi}$ is a proper function of the phase-space point $\Phi_n=\gbar_i(x)$, and not of the latent variable $x$ of channel $i$ to ensure correct normalization.
Because the CWnet often concentrates the channel weights in a small number of channels, we drop the smallest integration channels when their cumulative contribution falls below a given threshold.

\subsubsection*{Neural importance sampling}

The second step in the chain, from the unit hypercube $x$ to the latent space $z$, is an additional trainable transformation that further flattens the integrand. 
Classically it is performed by \vegas~\cite{Lepage:1977sw,Lepage:1980dq,Lepage:2020tgj}, which adapts a factorized grid and therefore cannot capture correlations between the integration variables. 
\madnis instead uses a more expressive normalizing flow~\cite{inn,Kobyzev_2020}
\begin{align}
\Phi_n
\xleftrightarrow[\hspace{1.5cm}]{\text{analytic}}
x \in [0,1]^D
\xleftrightarrow[\quad \leftarrow \overline{H}_{i,\theta}(z)\quad]{H_{i,\theta}(x)\rightarrow}
z \in [0,1]^D \; ,
\label{eq:flow}
\end{align}
with parameters $\theta$ and associated normalized Jacobian density
\begin{align}
   h_{i,\theta}(x) = \left\vert\frac{\partial H_{i,\theta}(x)}{\partial x}\right\vert
   \qquad \mwith \qquad \int \d x\,h_{i,\theta}(x) = 1 \; .
   \label{eq:madnis-density}
\end{align}
Events are generated by drawing $z$ uniformly and inverting the chain, $x=\overline H_{i,\theta}(z)$ followed by $\Phi_n=\gbar_i(x)$, so that $x$ now follows $h_{i,\theta}$ rather than the uniform distribution of \cref{eq:mc-master-x}. 
In terms of the physical phase space this corresponds to refining the analytic channel density of \cref{eq:channel-density} into the trained density
\begin{align}
g_i(\Phi_n)
\;\longrightarrow\;
g_{i,\theta}(\Phi_n)
= h_{i,\theta}\bigl(G_i(\Phi_n)\bigr)\; g_i(\Phi_n) \; ,
\label{eq:flow-density}
\end{align}
which makes the importance-sampling interpretation explicit, but is never evaluated in this form. 
Instead, we only need the ratio following directly from \cref{eq:fbar},
\begin{align}
  \frac{f(\Phi_n)}{g_{i,\theta}(\Phi_n)} = \frac{\bar f_i(x)}{h_{i,\theta}(x)} \; ,
  \label{eq:ratio-x}
\end{align}
\ie we handle all densities on the hypercube.
Since \vegas converges quickly for factorizing integrands, we employ a \vegas pre-training as a fast and stable starting point for the flow~\cite{Heimel:2023ngj}.

\subsubsection*{Variance loss}

With both networks in place, \cref{eq:mc-master-x} becomes
\begin{align}
  I[f]=\sum_{i=1}^{N_c}\left\langle
  \frac{\alpha_{i,\xi}\!\left(\gbar_i(x)\right)\, \bar f_i(x)}{h_{i,\theta}(x)}
  \right\rangle_{x\sim h_{i,\theta}(x)}\;,
\end{align}
and we train \madnis by minimizing the total variance.
Encoding the optimal, stratified distribution of points across channels~\cite{Press:1989vk,Weinzierl:2000wd} gives the multi-channel loss
\begin{align}
\loss_\madnis
= \left[\sum_{i=1}^{N_c}\left(
  \left\langle \frac{\alpha_{i,\xi}\!\left(\gbar_i(x)\right)^2  \bar f_i(x)^2}{h_{i,\theta}(x)\,q_i(x)} \right\rangle_{x\sim q_i(x)}
  - \left\langle \frac{\alpha_{i,\xi}\!\left(\gbar_i(x)\right)\, \bar f_i(x)}{q_i(x)} \right\rangle_{x\sim q_i(x)}^2
  \right)^{1/2}\right]^2 ,
\label{eq:madnis-loss}
\end{align}
where $q_i(x)$ is the sampling density of the training sample. Its relation to the trained density $h_{i,\theta}(x)$ depends on the training mode and is specified below.

\subsubsection*{Online and buffered training}

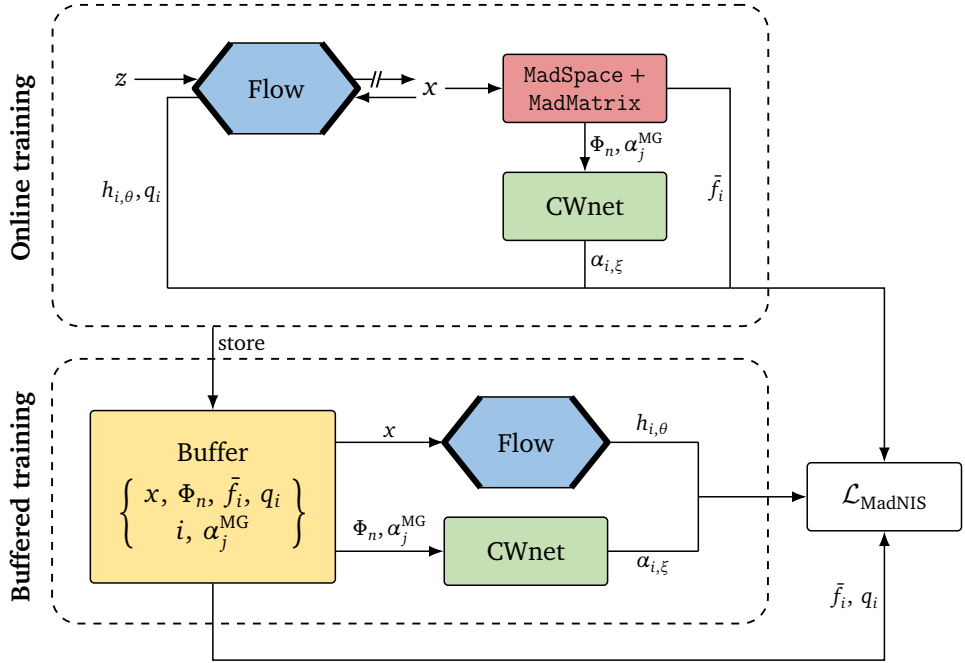
\begin{figure}[t]
\centering
\definecolor{Rcolor}{HTML}{E99595}
\definecolor{Gcolor}{HTML}{C5E0B4}
\definecolor{Bcolor}{HTML}{9DC3E6}
\definecolor{Ycolor}{HTML}{FFE699}

\tikzset{
  txt/.style   = {align=center, font=\LARGE},
  box/.style   = {rectangle, align=center, rounded corners=0.3ex, font=\Large,
                  minimum width=3.6cm, minimum height=1.5cm, fill=white, draw,
                  line width=0.2mm},
  cwnet/.style = {box, fill=Gcolor, font=\LARGE},
  arrow/.style = {thick,-{Latex[scale=1.0]}, line width=0.2mm, color=black},
  line/.style  = {thick, line width=0.2mm, color=black},
  lbl/.style   = {font=\Large, inner sep=3pt},
  side/.style  = {font=\LARGE, rotate=90},
  frame/.style = {draw, dashed, line width=0.25mm, rounded corners=0.35cm,
                  inner sep=0.5cm}
}

\newcommand{\flowhex}[3]{%
  \begin{scope}[shift={#2}]
    \fill[Bcolor] (-1.8,0) -- (-1.0,1.0) -- (1.0,1.0) -- (1.8,0)
               -- (1.0,-1.0) -- (-1.0,-1.0) -- cycle;
    \draw[line width=0.2mm] (-1.8,0) -- (-1.0,1.0) -- (1.0,1.0) -- (1.8,0)
               -- (1.0,-1.0) -- (-1.0,-1.0) -- cycle;
    \draw[line width=0.8mm, line join=round] (-1.0,1.0) -- (-1.8,0) -- (-1.0,-1.0);
    \draw[line width=0.8mm, line join=round] ( 1.0,1.0) -- ( 1.8,0) -- ( 1.0,-1.0);
    \node[align=center, font=\LARGE] at (0,0) {#3};
    \coordinate (#1-wu) at (-1.72, 0.20);
    \coordinate (#1-wl) at (-1.70,-0.20);
    \coordinate (#1-eu) at ( 1.70, 0.20);
    \coordinate (#1-el) at ( 1.72,-0.20);
  \end{scope}
  \node (#1) [inner sep=0pt, minimum width=3.6cm, minimum height=2.0cm] at #2 {};}

\newcommand{\cut}[1]{%
  \draw[white, line width=0.3mm]
    ($#1+(-0.08,0)$) -- ($#1+(0.08,0)$);
  \draw[line width=0.25mm]
    ($#1+(-0.11,-0.18)$) -- ($#1+(-0.01,0.18)$);
  \draw[line width=0.25mm]
    ($#1+( 0.01,-0.18)$) -- ($#1+( 0.11,0.18)$);
}

\begin{tikzpicture}[scale=0.6, every node/.style={transform shape}]

\node (z) [txt] at (0.0,1.2) {$z$};
\flowhex{fonflow}{(3.4,1.0)}{Flow}
\node (x) [txt] at (6.8,1.0) {$x$};
\node (mg) [box, fill=Rcolor] at (10.2,1.0) {\madspace $+$\\\madmatrix};
\node (cw) [cwnet] at (10.2,-1.6) {CWnet};
\node (busA) [inner sep=0pt, minimum size=0pt] at (13.4,-3.4) {};
\node (busL) [inner sep=0pt, minimum size=0pt] at ( 1.1,-3.4) {};

\node (bu) [box, fill=Ycolor, minimum width=5.4cm, minimum height=3.8cm, font=\LARGE]
  at (2.0,-8.0)
  {Buffer\\[0.3cm]
   $\left\{\begin{array}{c}
      x,\;\Phi_n,\;\bar f_i,\;q_i\\[0.05cm]
      i,\;\alpha^{\mathrm{MG}}_j
    \end{array}\right\}$};
\flowhex{fbflow}{(8.9,-6.8)}{Flow}
\node (cb) [cwnet] at (8.9,-9.2) {CWnet};
\node (busB) [inner sep=0pt, minimum size=0pt] at (12.7,-8.0) {};

\node (ls) [box, minimum width=3.4cm,font=\LARGE] at (16.8,-8.0) {$\mathcal L_{\mathrm{MadNIS}}$};

\coordinate (fonL) at (-0.7,-1.0);
\coordinate (fonR) at (13.4,-1.0);

\coordinate (fbfL) at (-0.7,-8.0);
\coordinate (fbfR) at (13.4,-8.0);

\node (fon) [frame,
  fit=(z)(fonflow)(x)(mg)(cw)(busA)(busL)(fonL)(fonR)] {};

\node (fbf) [frame,
  fit=(bu)(fbflow)(cb)(busB)(fbfL)(fbfR)] {};

\node [side] at (-2.2,-0.75) {\textbf{Online training}};
\node [side] at (-2.2,-8.0) {\textbf{Buffered training}};

\coordinate (xup) at ($(x.west)+(0,0.20)$);
\coordinate (xlo) at ($(x.west)+(0,-0.20)$);

\draw[arrow] (z.east) -- (fonflow-wu);
\draw[arrow] (fonflow-eu) -- (xup);
\draw[arrow] (x.east) -- (mg.west);
\draw[arrow] (xlo) -- (fonflow-el);
\draw[line]  (fonflow-wl) -- (1.0,0.8) -- (1.0,-3.4)
  node[lbl, pos=0.5, left] {$h_{i,\theta}, q_i$};
\draw[arrow] (mg.south) -- (cw.north) node[lbl, midway, right] {$\Phi_n, \alpha^{\mathrm{MG}}_j$};
\draw[line]  (mg.east) -- (13.4,1.0) -- (13.4,-3.4)
  node[lbl, pos=0.5, left] {$\bar f_i$};
\draw[line]  (cw.south) -- (10.2,-3.4) node[lbl, midway, right] {$\alpha_{i,\xi}$};
\draw[line]  (1.0,-3.4) -- (13.4,-3.4);
\cut{(5.6,1.20)}

\draw[arrow] (bu.east |- fbflow) -- (fbflow.west) node[lbl, pos=0.5, above] {$x$};
\draw[arrow] (bu.east |- cb)     -- (cb.west)     node[lbl, pos=0.5, above] {$\Phi_n, \alpha^{\mathrm{MG}}_j$};
\draw[line]  (fbflow.east) -- (12.7,-6.8) node[lbl, midway, above] {$h_{i,\theta}$};
\draw[line]  (cb.east)     -- (12.7,-9.2) node[lbl, midway, below] {$\alpha_{i,\xi}$};
\draw[line]  (12.7,-6.8) -- (12.7,-9.2);

\draw[arrow] (busA) -| (ls.north);
\draw[arrow] (busB) -- (ls.west);
\draw[arrow] (bu.south) -- (2.0,-11.6) -| (ls.south)
  node[lbl, pos=0.75, left] {$\bar f_i,\;q_i$};
\draw[arrow] (fon.south -| bu) -- (bu.north) node[lbl, pos=0.2, right] {store};

\end{tikzpicture}
\caption{Online and buffered training in \madnis. Both branches minimize the same loss, \cref{eq:madnis-loss}, evaluated with the proposal $q_i$ of \cref{eq:proposal}. The break in the sampling arrow marks where the gradient is detached, as the generation pass $z \to x$ carries no gradients, so the flow is evaluated a second time in the forward direction to provide $h_{i,\theta}$ as a proper function of $x$.
Both networks are evaluated in either branch, so only the online branch calls \madspace and \madmatrix.}
\label{fig:buffered_training}
\end{figure}

The loss of \cref{eq:madnis-loss} is evaluated on points drawn from a proposal $q_i$ which is deliberately kept distinct from the trained density $h_{i,\theta}$.
Since $q_i$ appears explicitly in the loss, any density that is non-zero wherever the integrand contributes gives an unbiased estimate of the same variance, and a mixture of several densities can for instance stabilize the training~\cite{Heimel:2026cxh}.
We use two choices, which define two training modes~\cite{Heimel:2022wyj,Heimel:2023ngj},
\begin{align}
  q_i(x) = h_{i,\theta^*}(x)\big\vert_\text{no-grad},
  \qquad \mwith \qquad
  \theta^* =
  \begin{cases}
    \theta                & \text{online} \\
    \theta_\text{gen}     & \text{buffered} \; .
  \end{cases}
  \label{eq:proposal}
\end{align}
In online training, the proposal $q_i$ is evaluated with the current parameters $\theta$ and is therefore numerically identical to $h_{i,\theta}$, but detached from the gradient computation.
In buffered training, $\theta_\text{gen}$ denotes the network parameters at the time the point was generated. 
Both modes therefore estimate the same variance and can be interleaved freely.
The two branches are illustrated in \cref{fig:buffered_training} and described below:
\begin{description}
\item[Online training:] We draw $z$ uniformly and invert the flow, $x=\overline H_{i,\theta}(z)$, then call \madspace and \madmatrix once to obtain the phase-space point $\Phi_n=\gbar_i(x)$, the integrand $\bar f_i(x)$, and the channel-weight priors $\alpha^\text{MG}_j(\Phi_n)$.
This sampling pass has to run without gradients, as differentiating through it would require $\nabla_x \bar f_i(x)$, which is not available for a generic matrix element~\cite{Heimel:2024wph}.
We therefore evaluate the density in a second, forward pass $z=H_{i,\theta}(x)$, so that $h_{i,\theta}$ enters the loss as a proper function of $x$~\cite{Heimel:2022wyj}, and the channel weights in a pass of the CWnet over $\Phi_n$.
An online update thus costs one integrand evaluation and two flow passes.

\item[Buffered training:] We reuse points generated during earlier online steps and only perform the weight update.
Of the ingredients of \cref{eq:madnis-loss}, we read $\bar f_i(x)$ and $q_i(x)$ from the buffer, while we recompute $h_{i,\theta}(x)$ and $\alpha_{i,\xi}(\Phi_n)$ with the current parameters.
We therefore store the hypercube point $x$ for the flow, the phase-space point $\Phi_n$ for the CWnet, the integrand $\bar f_i(x)$, the sampling density $q_i(x)$, the generating channel index $i$, and the priors $\alpha^\text{MG}_j$.
A buffered update thus costs a single flow pass and no integrand evaluation, saving the \madspace call and the costly matrix-element evaluation in \madmatrix.
\end{description}

The multi-channel weight priors dominate the memory footprint of the buffer and therefore limit the number of points it can hold.
The softmax normalization of \cref{eq:alpha-net} couples all $N_c$ channels, so in principle the complete set $\{\alpha^\text{MG}_j\}$ is required for every buffered point, which amounts to several gigabytes for $\mathcal{O}(10^6)$ points and $\mathcal{O}(10^3)$ channels. 
In practice only a few channels contribute significantly to any given point, so we store the index-weight pairs $(j,\alpha^\text{MG}_j)$ of the $m \ll N_c$ largest contributions, always including the channel the point was generated in, and set the remaining weights to zero.
We adjust the normalization accordingly to leave the result unbiased.

\clearpage
\section{Improving robustness and scalability}
\label{sec:technical}

Previous \madnis studies~\cite{Heimel:2023ngj} focused on the achievable gain in unweighting efficiency for single partonic subprocesses.
There, a common set of hyperparameters was sufficient, and it did not matter if a network was larger or a training longer than necessary.
As the default in \mg, \madnis instead has to handle full proton-proton processes with many partonic subprocesses, run unattended across a wide range of processes and multiplicities, and stay efficient for the moderate event counts users typically request, where the training cost directly adds to the total cost.
We therefore extend the method to discrete degrees of freedom, reduce the number of independent trainings, make buffered training more robust, and speed up the training itself, before we choose the hyperparameters through automated defaults that balance expressivity against throughput and training time.
Most of these improvements also pay off when an expert tunes the setup.
Combined with \madspace and \madmatrix, \madnis now provides complete event generation, which unweights the events and writes them out with their flavors, helicities, and color flows in addition to their four-momenta.
The whole chain from random-number generation to unweighted events thereby runs on the GPU.


\subsection{Sampling discrete degrees of freedom}
\label{sec:discrete-sampling}

The parton-level integrand generally depends on both the four-momenta $\Phi_n$ and on discrete degrees of freedom, such as the helicity, color, and flavor configuration.
On top of these physical quantum numbers, the phase-space construction introduces a further discrete label, which picks one of several equivalent orderings of the final-state momenta inside an integration channel.
Whenever the integrand depends on such a label, an efficient sampler has to account for it, so in general we condition the sampling on all of them.

In default \mg, however, we sum over helicities and colors, so this dependence drops out of the cross section.
We only assign a specific helicity and color configuration when we write out events, but this assignment leaves the integral unchanged.
Other generators do sample these quantum numbers, for instance through the color-sampled recursion of \texttt{Comix} in \sherpa~\cite{Gleisberg:2008fv}, the leading-color event generation with subsequent full-color reweighting of \amplicol~\cite{Frederix:2024uvy,Frederix:2026ejl}, or the helicity-conditioned normalizing flows combined with \pepper~\cite{Bothmann:2026yzp}.
Our setup can accommodate these cases, as we discuss at the end of this section.
For \mg, we are therefore left with the symmetry and flavor configurations:
\begin{itemize}
\item \textbf{Symmetry configuration.}
It arises from the phase-space construction with single-diagram enhancement, in which Feynman diagrams with the same phase-space topology but different momentum orderings of the final-state particles are grouped into a single integration channel with the same phase-space mapping.
After evaluating the mapping, we have to pick one specific permutation and apply it to the momenta.
If only diagrams with the same physical structure, including propagator types and masses, are grouped, this is an exact symmetry and uniform sampling of the permutation is sufficient.
More flexible channel grouping, as introduced in \cref{sec:merging}, or the trainable channel weights from \madnis break this symmetry, so adaptive importance sampling of the momentum permutations reduces the variance of the phase-space integral.

\item \textbf{Flavor configuration.}
\mg merges multiple partonic processes into a single cross-section integral when they share similar physics, and determines the specific flavor configuration via importance sampling.
The main difference among flavor configurations comes from the parton-density functions (PDFs), which depend only on the initial-state flavors.
Matrix elements, in contrast, may depend on all flavors. They differ either through flavor-dependent couplings within a given diagram, for instance for the $W$ boson in electroweak processes, or through the number of contributing diagrams.
For example, $\Pu \Pubar \to \Pu \Pubar$ receives both $s$- and $t$-channel contributions, whereas $\Pu \Pdbar \to \Pu \Pdbar$ has only $t$-channel diagrams and therefore fewer of them.
\end{itemize}
We combine continuous and discrete degrees of freedom using an autoregressive decomposition of the full sampling probability, $h_\theta(x, r, a)$, with indices $r$ and $a$ enumerating the symmetry and flavor configurations available in a given integration channel, respectively.
We construct this sampling for each integration channel separately and omit the channel index $i$ in the following for readability.
For \madnis, we decompose the probability as
\begin{equation}
    h_\theta(x, r, a) = h_\theta(x) \, h_\theta(r \vert x) \, h_\theta(a \vert x, r) \; .
    \label{eq:autoregressive}
\end{equation}
This results in a more expressive architecture compared to a setup with a flow conditioned on the discrete indices because the spline-coupling-based normalizing flows used for the continuous dimensions have a more restrictive architecture than the simple MLP networks predicting the discrete probabilities.
The flows used with \pepper follow this conditional approach for the helicity configuration~\cite{Bothmann:2026yzp}.
For the symmetry configurations, we use a fully connected network $S_\theta(x)$ that predicts a vector of unnormalized log-probabilities. We then compute the probability $h_\theta(r \vert x)$ using a softmax function,
\begin{equation}
h_\theta(r \vert x) = \frac{
    \exp S_{r,\theta}(x)
}{
    \sum_{r'} \exp S_{r',\theta}(x)
} \; .
\label{eq:symm-prob}
\end{equation}
For the flavor sampling we exploit that the PDFs dominate the flavor dependence of the differential cross section and are cheap to evaluate. Having sampled $x$ and $r$ first, we can evaluate the phase-space point $\Phi_n \equiv \Phi_n(x, r)$ to obtain the momentum fractions $x_1$ and $x_2$, as well as the factorization scale $\mu_\mathrm{F}$. Using $a_1$ and $a_2$ to denote the flavors of the two incoming partons for the flavor configuration $a$, this defines the unnormalized weight
\begin{equation}
    w_\text{PDF}(a \vert \Phi_n) =
    \text{PDF}_{a_1}(x_1, \mu_\mathrm{F}) \:
    \text{PDF}_{a_2}(x_2, \mu_\mathrm{F}) \; .
    \label{eq:pdf-weight}
\end{equation}
Not every flavor configuration contributes for a given symmetry configuration, because certain combinations have no physical diagram, so their channel weight vanishes and they never contribute to the integral. We therefore restrict the flavor sampling to the set
\begin{align}
  \mathcal{F}_r = \left\{ a 
  \mid \text{flavor configuration $a$ has a contributing diagram for permutation $r$} \right\} \; ,
  \label{eq:allowed-set}
\end{align}
and normalize the weights of \cref{eq:pdf-weight} over this set,
\begin{equation}
h_\text{PDF}(a \vert x, r) = \frac{
    w_\text{PDF}(a \vert \Phi_n(x, r))
}{
    \sum_{a'  \in \mathcal{F}_r} w_\text{PDF}(a' \vert \Phi_n(x, r))
}
\qquad \mand \qquad
h_\text{PDF}(a \vert x, r) = 0 \quad \text{for } a \notin \mathcal{F}_r \; .
\label{eq:pdf-prior}
\end{equation}
For processes in which the matrix element does not depend on the flavor configuration, this prior distribution already yields the perfect importance-sampling distribution. To also account for the flavor dependence of the matrix element, we apply a learned multiplicative correction, exactly as for the \mg channel weights in \cref{eq:alpha-correction},
\begin{equation}
h_\theta(a \vert x, r) = \frac{
    h_\text{PDF}(a \vert x, r) \,\exp F_{a,\theta}(x, r)
}{
    \sum_{a'  \in \mathcal{F}_r} h_\text{PDF}(a' \vert x, r) \,\exp F_{a',\theta}(x, r)
} \; ,
\label{eq:flav-prob}
\end{equation}
where $F_\theta(x, r)$ is a second fully connected network with an output for each flavor configuration.
Our discrete sampling setup extends directly to more discrete dimensions, for example to perform helicity sampling instead of summing, or to sample FKS sectors in the NLO extension of \madnis~\cite{DeCrescenzo:2026tsp}.

\subsection{Merging matrix elements, channels and subprocesses}
\label{sec:merging}

\mg organizes the phase-space integration and event generation of a full process into a two-level hierarchy.
\begin{itemize}
    \item A \emph{subprocess} groups one or more partonic flavor configurations, where in the latter case flavor sampling as introduced in \cref{sec:discrete-sampling} is applied. In \mgfive a subprocess could bundle multiple independent matrix-element implementations whereas in \madmatrix there is a one-to-one correspondence between matrix element code and subprocesses.
    \item Each subprocess is divided into \emph{integration channels} using the channel weights introduced in \cref{eq:mg-prior}. There is an independent \madnis network or \vegas grid for each channel. The integral remains unbiased as long as these channel weights are normalized to one at every phase-space point. Consequently there is a lot of freedom in the way the subprocess is decomposed into channels.
\end{itemize}
The construction of subprocesses and channels is a trade-off between many factors, including \madnis and \vegas training time, complexity of each individual integrand, memory usage and code size. As \vegas integration grids converge relatively fast while only providing limited expressivity, event generators like \mgfive or \sherpa use a large number of integration channels. However, even in a \vegas setup, building a large number of integration channels leads to diminishing returns or even starts to complicate the phase-space integration going towards high-multiplicity processes. Neural importance sampling offers higher expressiveness at the cost of much longer training times. Therefore, it profits less from a large amount of integration channels. Indeed, we find that often a large fraction of the integration channels is dropped during the multi-channel \madnis training~\cite{Heimel:2023ngj}. Even with channel dropping, a large number of channels still leads to considerable computational cost and decreased training stability, especially during the early stages of the training when most channels are still active. Consequently, the \madnis training can be improved by building fewer channels in the first place.

\begin{figure}[t]
\centering
\input{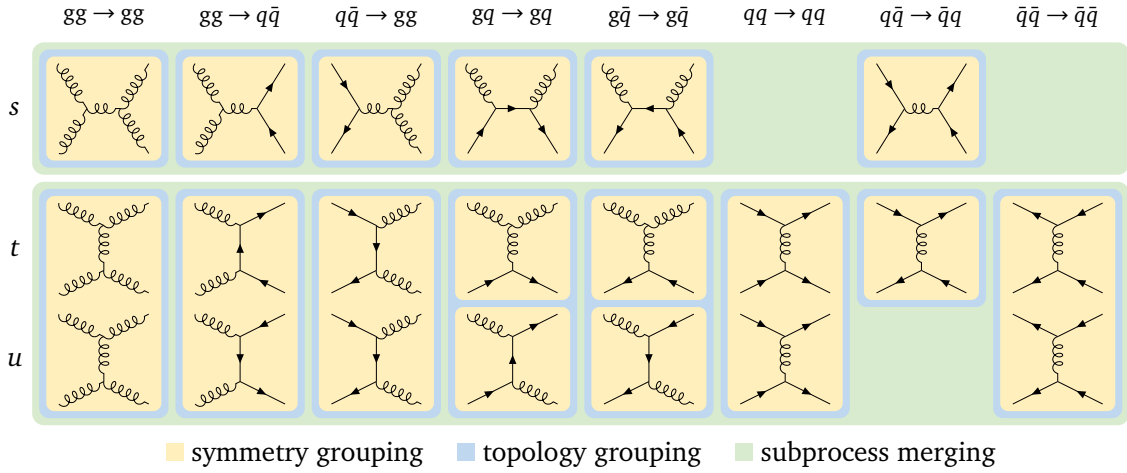}
\caption{Illustration of symmetry grouping as in \mgfive, topology grouping and subprocess merging for the Feynman diagrams contributing to $\Pp \Pp \to \Pj \Pj$. Columns correspond to the eight subprocesses with an independent matrix-element implementation. Rows correspond to the $s$-, $t$-, and $u$-channel topologies. We omit the diagram for $\Pg \Pg \to \Pg \Pg$ with a quartic gluon coupling because it is not explicitly mapped out by a phase-space mapping.}
\label{fig:subproc-merging}
\end{figure}

Orthogonal to these improvements, GPU-based or SIMD-vectorized setups favor processing samples in large batches with as little divergence between code paths as possible. Therefore, they profit from merging the matrix-element implementations for related partonic channels. To tackle these various challenges, we implement the following improvements:
\begin{description}
\item[Merging matrix-element implementations:]
A subprocess can contain several different matrix elements. For instance, the process $\Pu \Pd \to \Pu \Pd$ has fewer Feynman diagrams compared to $\Pu \Pu \to \Pu \Pu$. In \mgfive, these processes are grouped into the same subprocess, but have independent matrix-element codes that are called depending on the flavor index. However, this is not optimal for a GPU- or SIMD-vectorized implementation as it would require subdividing each batch of points. Instead, we merge all code paths in a subprocess and decide at the level of individual couplings whether they are present or zero for the current flavor configuration.

\item[Merging integration channels of similar topology:]
In many cases, Feynman diagrams share the same initial- and final-state particles and the same diagram topology but differ in the type of propagator. For instance, as illustrated in \cref{fig:subproc-merging}, $\Pg \Pq \to \Pg \Pq$ has a $t$-channel diagram with a gluon propagator and a $u$-channel diagram with a quark propagator. Although both are massless propagators, their amplitudes differ, and \mgfive therefore builds independent channels for them. However, because they share the same topology and propagator masses, these channels will have the same analytic phase-space mapping. Instead, we group these diagrams into the same channel and address the differences through the learnable symmetry sampling introduced in \cref{sec:discrete-sampling}.

\item[Merging subprocesses:]
By default, we group partonic processes according to whether each external parton is a quark, an antiquark, or a gluon. This is similar to the grouping in the \pepper event generator that was studied in Ref.~\cite{Bothmann:2026yzp}.
Going one step further, we can also merge diagrams of the same topology across subprocesses. 
For the example of $\Pp \Pp \to \Pj \Pj$, this results in only two integration channels, denoted as green boxes in \cref{fig:subproc-merging}: one for the $s$-channel and one for the $t$- and $u$-channel diagrams. 
The subprocess for a given event is then chosen at the learnable flavor-sampling level. 
While there can be a large variation in physics distributions for different subprocesses because of the large differences between the quark and gluon PDFs, most of this is addressed using the PDF prior imposed during the flavor sampling, see \cref{sec:discrete-sampling}, so it does not have to be learned by a neural network. 
However, the training task after merging subprocesses is still more complicated than per-subprocess training, and evaluating the matrix element requires subdividing the batch, which results in smaller batch sizes and potentially lower GPU throughput. 
For this reason, we make subprocess merging optional, and finding smart heuristics to decide when to enable it remains a challenge to be investigated in the future.

\item[Further reduction of channel count:]
Even with all the improvements discussed above, the channel count can still be excessive for some processes. For example, some subprocesses of $\Pp \Pp \to \Pt \Ptbar + 3\Pj$ have 120 integration channels. We find that for massless QCD interactions, $t$-channel mappings combined with expressive \madnis sampling are usually sufficient to model the integrand features precisely. Consequently, we discard channels with $s$-channel propagators if the number of channels would otherwise exceed a threshold of 16, and adjust the channel-weight normalization to keep the result unbiased. For $\Pt \Ptbar + 3\Pj$, this reduces the number of channels to twelve, resulting in significantly faster and more stable \madnis training. Additionally, \madspace offers several alternative mappings, such as \texttt{RamboOnDiet} or \texttt{Chili}, that can be used as a single integration channel instead of the classic multi-channel approach.

\end{description}

\subsection{Buffered training on unweighted events}
\label{sec:buffered-unweighted}

Buffered training makes \madnis training more efficient by reusing samples multiple times for optimization before replacing them with newly generated samples. In our previous work~\cite{Heimel:2023ngj}, we implemented this by storing all weighted events generated during online training that pass phase-space cuts. While this works well for simple processes, we find that for more complex tasks, it can lead to training instability and overfitting compared to pure online training. Moreover, as the effective sample size decreases with process complexity, less effective statistics can be stored with the same amount of memory. Ref.~\cite{midgley2022flow} proposes a prioritized replay buffer as a solution to the first problem. Training batches are sampled from the buffer with probabilities proportional to their event weights. While this improves training stability by increasing the effective sample size per batch, it does not solve the issue of reduced statistics across the entire buffer. A related sample replay supplements the online training of the \pepper flows~\cite{Bothmann:2026yzp}.

We address both problems by performing a partial unweighting step before storing events into the buffer. To this end, for a batch of $K$ points $x_k$ generated in channel $i$ during online training, we compute the event weights
\begin{equation}
    w_k = \frac{\alpha_{i,\xi}(\Phi_{n,k})\,\bar f_i(x_k)}{q_i(x_k)}
    \qquad \mwith \qquad
    \Phi_{n,k} = \gbar_i(x_k) \; ,
    \label{eq:buffer-weight}
\end{equation}
and use them to determine a batch-wise unweighting threshold $w_\text{max}$ as a quantile of the $w_k$. Next, we draw uniformly distributed $u_k \in [0,1]$ and perform a standard accept-reject step,
\begin{equation}
    \tilde{w}_k =
    \begin{cases}
        \max(w_k, w_\text{max}) & \text{if } w_k \ge u_k\, w_\text{max} \\
        0 & \text{otherwise}
    \end{cases} \; .
    \label{eq:partial-unweighting}
\end{equation}
For a quantile of $0.95$, this means that between $5\%$ and $100\%$ of the points are accepted, depending on how narrow the weight distribution is.
Points failing the phase-space cuts have $w_k = 0$ and are therefore always rejected.

Both training modes only use a subset of the $K$ generated points.
Online training uses the $K_\text{cut}$ points that pass the phase-space cuts, while buffered training only stores the $K_\text{acc}$ points accepted by \cref{eq:partial-unweighting}.
To preserve the property that $\langle w_k \rangle$ yields the integral of the integration channel, we rescale the sampling densities of the remaining points as
\begin{equation}
    \tilde{q}_i(x_k) = q_i(x_k) \times
    \begin{cases}
        \dfrac{K}{K_\text{cut}} & \text{online} \\[2ex]
        \dfrac{w_k}{\tilde{w}_k}\, \dfrac{K}{K_\text{acc}} & \text{buffered}
    \end{cases} \; ,
    \label{eq:rescale-q}
\end{equation}
and use $\tilde q_i(x_k)$ in place of $q_i(x_k)$ in the loss.
Online and buffered training batches then remain compatible with the multi-channel variance loss of \cref{eq:madnis-loss}, and their loss values stay consistent.

Even with partially unweighted points, buffered training can still overfit and reduce the training performance.
We prevent this by choosing the buffered-training hyperparameters such that all points in the buffer are replaced frequently, and the weight distribution of the buffered points is stable enough not to destabilize the training.
We do not store points from the first 1000 training iterations, as both the flow and the channel weights change too quickly in this initial phase.
We start training on buffered batches only once the buffer holds $10^4$ points, and ramp up the fraction of buffered iterations, linearly as the buffer fills.
We limit the buffer to $6 \times 10^4$ points per channel and set the quantile for $w_\text{max}$ to $0.95$, so the buffer content is replaced often enough to prevent overfitting.

\subsection{Concurrent training and generation}
\label{sec:framework}

In event generation with neural importance sampling, the trade-off between training time and unweighting efficiency is a key consideration for achieving overall speed-up. It is therefore crucial not only to optimize neural network architecture and training hyperparameters, but also to minimize the computational cost per training iteration.

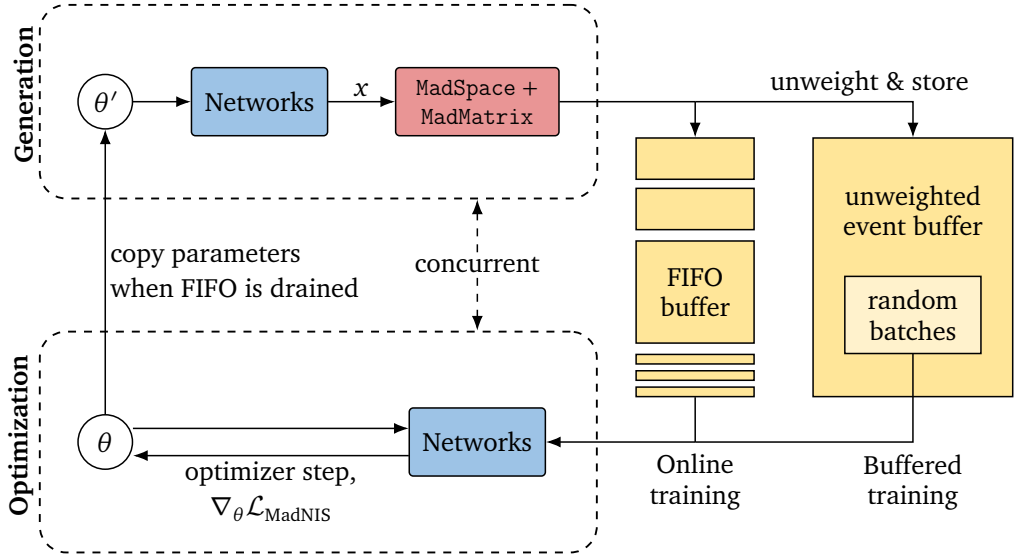
\begin{figure}[t]
\centering
\tikzset{
  box/.style      = {rectangle, align=center, rounded corners=0.3ex, font=\Large,
                     minimum width=3.6cm, minimum height=1.5cm, fill=white, draw,
                     line width=0.2mm},
  nets/.style     = {box, fill=Bcolor, font=\LARGE, minimum width=3.0cm},
  par/.style      = {circle, draw, line width=0.2mm, fill=white, font=\LARGE,
                     minimum size=1.2cm, inner sep=0pt},
  arrow/.style    = {thick,-{Latex[scale=1.0]}, line width=0.2mm, color=black},
  line/.style     = {thick, line width=0.2mm, color=black},
  lbl/.style      = {font=\LARGE, inner sep=3pt},
  side/.style     = {font=\LARGE, rotate=90},
  frame/.style    = {draw, dashed, line width=0.25mm, rounded corners=0.35cm,
                     inner sep=0.5cm},
  cell/.style     = {draw, line width=0.2mm, fill=Ycolor}
}

\newcommand{\fwbox}[5]{%
  \draw[cell] ({#2-#4/2},#3) rectangle ({#2+#4/2},{#3-#5});
  \node (#1) [inner sep=0pt, fit={({#2-#4/2},#3) ({#2+#4/2},{#3-#5})}] {};}

\begin{tikzpicture}[scale=0.6, every node/.style={transform shape}]
\def\fw{2.6}   
\def\bw{4.4}   
\def\lh{0.91}  
\def\sh{0.21}  

\node (gpar) [par] at (1.2,0.0) {$\theta'$};
\node (gnet) [nets] at (4.6,0.0) {Networks};
\node (mg) [box, fill=Rcolor] at (9.4,0.0) {\madspace $+$\\\madmatrix};

\fwbox{lb1}{14.2}{-0.8}{\fw}{\lh}
\fwbox{lb2}{14.2}{-1.91}{\fw}{\lh}
\fwbox{fifo}{14.2}{-3.07}{\fw}{2.25}
\fwbox{fs1}{14.2}{-5.57}{\fw}{\sh}
\fwbox{fs2}{14.2}{-5.93}{\fw}{\sh}
\fwbox{fs3}{14.2}{-6.29}{\fw}{\sh}
\node [lbl, align=center] at (fifo.center) {FIFO\\[-0.1cm]buffer};

\fwbox{buf}{19.0}{-0.8}{\bw}{5.7}
\node [lbl, align=center] at (19.0,-2.4) {unweighted\\[-0.1cm]event buffer};
\node (rnd) [cell, fill=Ycolor!50, font=\LARGE, align=center,
  minimum width=3.0cm, minimum height=1.7cm] at (19.0,-4.7)
  {random\\[-0.1cm]batches};

\node (opar) [par] at (1.2,-7.5) {$\theta$};
\node (onet) [nets] at (9.4,-7.5) {Networks};

\coordinate (genT) at ( 1.2, 1.3);
\coordinate (genR) at (11.2,-1.3);
\coordinate (optT) at ( 1.2,-5.9);
\coordinate (optR) at (11.2,-9.1);

\node (fgen) [frame, fit=(gpar)(gnet)(mg)(genT)(genR)] {};
\node (fopt) [frame, fit=(opar)(onet)(optT)(optR)] {};

\node [side] at (fgen.west) [above=0.15cm] {\textbf{Generation}};
\node [side] at (fopt.west) [above=0.15cm] {\textbf{Optimization}};

\draw[arrow] (gpar.east) -- (gnet.west);
\draw[arrow] (gnet.east) -- (mg.west) node[lbl, midway, above] {$x$};

\draw[line] (mg.east) -- (lb1.north |- mg) coordinate (junc);
\draw[arrow] (junc) -- (lb1.north);
\draw[arrow] (junc) -| (buf.north)
  node[lbl, pos=0.4, above] {unweight \& store};

\coordinate (merge) at (fs3.south |- onet);
\draw[line] (fs3.south) -- (merge);
\draw[line] (rnd.south) |- (merge);
\draw[arrow] (merge) -- (onet.east);
\node [lbl, align=center, below=0.15cm] at (merge)
  {Online\\[-0.1cm]training};
\node [lbl, align=center, below=0.15cm] at (rnd.south |- merge)
  {Buffered\\[-0.1cm]training};

\draw[arrow] ($(opar.east)+(0,0.3)$) -- ($(onet.west)+(0,0.3)$);
\draw[arrow] ($(onet.west)+(0,-0.3)$) -- ($(opar.east)+(0,-0.3)$)
  node[lbl, midway, below, align=center] {optimizer step,\\$\nabla_\theta \mathcal L_{\mathrm{MadNIS}}$};

\draw[arrow] (opar.north) -- (gpar.south)
  node[lbl, midway, right, align=left]
  {copy parameters\\when FIFO is drained};

\node (conc) [lbl] at (9.4,0 |- {$(fgen.south)!0.5!(fopt.north)$}) {concurrent};
\draw[arrow, dashed] (conc.north) -- (conc.north |- fgen.south);
\draw[arrow, dashed] (conc.south) -- (conc.south |- fopt.north);

\end{tikzpicture}
\caption{Training workflow with decoupled generation and optimization. Networks denotes the normalizing flow, the channel-weight network, and the networks for the symmetry and flavor probabilities of \cref{sec:discrete-sampling}, and $\theta$ collectively denotes all their trainable parameters.}
\label{fig:framework}
\end{figure}

Especially when training on GPUs, the optimal batch size to maximize GPU throughput (typically tens of thousands) differs from the batch size that optimizes training performance by ensuring stable gradients while remaining data-efficient (typically hundreds to thousands). 
Consequently, a naive training loop that generates and optimizes training samples sequentially with the same batch size results in slow training and uses only a fraction of the GPU's capacity.
We solve this problem by decoupling training data generation and optimization, and running them concurrently. 
To this end, we keep two copies of the neural network in memory. We use one copy to optimize the network by consuming events in small batches, either from a small first-in-first-out (FIFO) buffer during online training or by randomly selecting samples from the larger buffer of unweighted events during buffered training. 
We use the second copy to independently generate events in large batches, thereby optimizing GPU throughput. 
The generation becomes active when the FIFO buffer is drained, at which point it copies the latest trainable parameters, generates weighted events, and stores them in the FIFO buffer. This setup preserves the short feedback loop between optimization and generated events in online training while allowing generation and optimization to run in parallel at their optimal batch sizes to maximize GPU or CPU usage. We illustrate our training workflow in \cref{fig:framework}.

Moreover, we include a fast implementation of RQ-spline flows~\cite{durkan2019neural} directly within the \madspace package that fuses the spline coupling blocks into a small number of compute kernels written in C++ and CUDA/HIP. They enable higher throughput than implementations in generic deep-learning frameworks such as \texttt{PyTorch} while removing overhead from interface and programming-language boundaries. This allows us to tightly interleave components such as normalizing flows, analytic phase-space mappings, and parton densities, making techniques like the PDF prior from \cref{sec:discrete-sampling} computationally efficient.

\subsection{Automated defaults}
\label{sec:defaults}
 
Selecting suitable hyperparameters for the \madnis networks and training is critical to ensure good training and event-generation performance. The best choice of parameters depends strongly on the process of interest: for simple, low-multiplicity processes, small networks suffice, and larger networks do not improve unweighting efficiency while reducing throughput because of the higher computational cost of evaluating the network. Moreover, for very simple processes and low event counts, training a neural importance sampler might not be amortized at all. For a production-ready event generator, it is not feasible to require an expert to tune these parameters for every new process. Instead, we define an automated procedure to make these decisions in a process-dependent way:
\begin{enumerate}
    \item Run a short \vegas survey with three iterations on 1000, 2000, and 4000 weighted events after cuts for each integration channel, similar to the survey in \mgfive.
    \item Use the relative standard deviation (RSD) after the third survey iteration, i.e., the relative integration error after removing the scaling with sample count, as a metric for the complexity of each subprocess. Determine the values of hyperparameters based on this metric.
    \item Use the \vegas grid from the survey pass either as a starting point for the full \vegas optimization, or for \vegas initialization of the \madnis flows.
    \item Run the \madnis or \vegas optimization.
\end{enumerate}
The most important hyperparameters to determine in a process-specific way are the width and depth of the flow sub-networks and the channel-weight network, to balance expressivity and throughput, and the batch size per integration channel, to balance training speed and convergence. Furthermore, we determine parameters such as the learning rate and the training iteration after which the learnable channel weights are frozen, as a function of the number of training iterations. Finally, choosing the training length is critical to achieve good unweighting efficiency while still amortizing training time, and we study it in detail in \cref{sec:training-time}. We show the full set of fixed and dynamic \madnis default parameters in \cref{tab:hyperparameters} in the appendix. While these defaults achieve good results across a wide range of physics processes, they still leave room for improvement and are likely to evolve.

\clearpage
\section{MadNIS in production}
\label{sec:production}

We now run \madnis as a user of \mgseven would, with its default settings and without any process-specific tuning.
We first validate the generated distributions against an independent baseline in \cref{sec:validation}, then quantify the training cost and its amortization in \cref{sec:training-time}, and finally benchmark a broad range of LHC processes in \cref{sec:process-zoo}.

All results use the default settings of \mgseven v0.2.1~\cite{MG7:v0.2.1} for proton-proton collisions at $\sqrt{s}=13$~TeV.
We use the NNPDF4.0 LO PDF set with $\alpha_s(\MZ)=0.118$~\cite{NNPDF:2021njg} provided by \lhapdf~\cite{Buckley:2014ana}, and dynamical factorization and renormalization scales $\mu_\text{F}=\mu_\text{R}=H_\text{T}/2$, where $H_\text{T}$ is the scalar sum of the transverse masses of all final-state particles.
The default generation cuts are
\begin{alignat}{3}
  p_{\text{T},j} &> 20~\text{GeV} \;, &\qquad
  p_{\text{T},\ell} &> 10~\text{GeV} \;, \qquad \notag\\
  |\eta_j| &< 5 \;, &\qquad
  |\eta_\ell| &< 2.5 \;, \qqquad
  \Delta R_{jj},\, \Delta R_{\ell\ell},\, \Delta R_{j\ell} > 0.4 \;,
  \label{eq:cuts}
\end{alignat}
where $\Pj$ denotes gluons and the four light quark flavors.
For the validation and the amortization study, we consider the processes
\begin{align}
  \Pp\Pp \to \Pt\Ptbar + n\Pj \quad \text{with} \quad n=1,2,3
  \qquad \mand \qquad
  \Pp\Pp \to \Pep\Pem + 2\Pj \; ,
  \label{eq:processes}
\end{align}
including all their allowed partonic subprocesses in the SM.

\subsection{Validation}
\label{sec:validation}

A faster sampler is only useful if it leaves the generated distributions unchanged.
We therefore compare differential distributions from \madnis with an independent \mgfive baseline for $\Pp\Pp \to \Pt\Ptbar + 2\Pj$ and $\Pp\Pp \to \Pep\Pem + 2\Pj$ from \cref{eq:processes}, following the validation of \madspace in Ref.~\cite{Heimel:2026hgp}.
We train \madnis for $10^4$ iterations with otherwise default settings from \mgseven, both with separate trainings for each subprocess and with a single merged training, as described in \cref{sec:merging}.
For each of them, we generate $10^6$ unweighted events while allowing for $0.1\%$ of the cross section to come from overweight events.
As baseline, we use \mgfive 3.8.1 with the same PDF set, scales, and cuts.
To ensure sufficient statistics for the baseline, denoted as \textit{MG5 sliced}, in both the bulk and the tails, we split each observable into ten slices and generate $10^6$ unweighted events in each of them.

\begin{figure}[t!]
\centering
\includegraphics[width=0.32\textwidth]{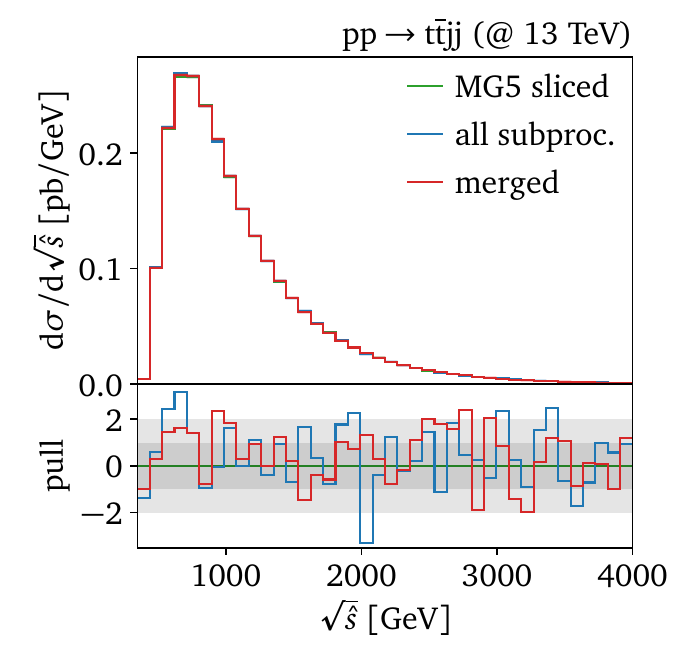}
\includegraphics[width=0.32\textwidth]{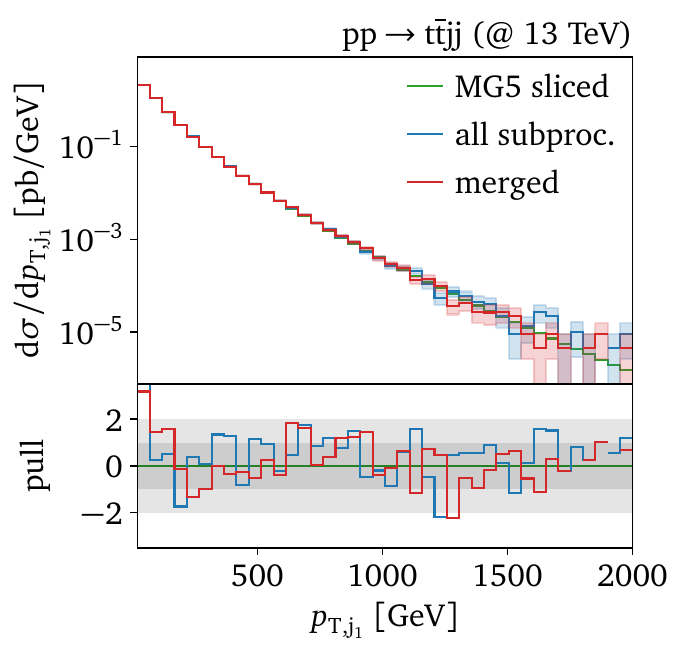}
\includegraphics[width=0.32\textwidth]{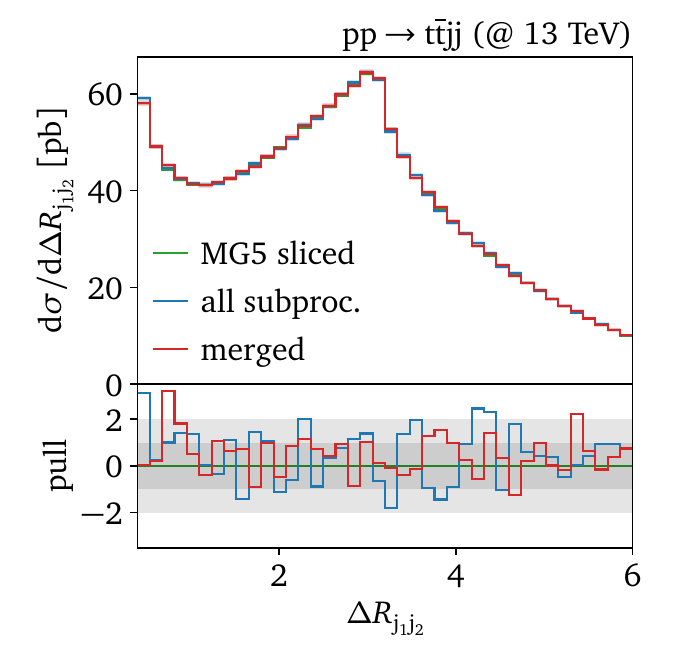}
\includegraphics[width=0.32\textwidth]{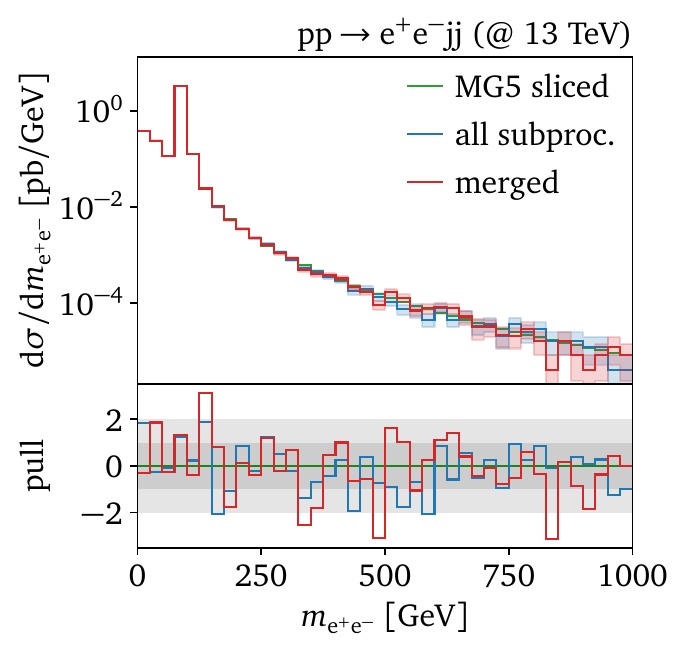}
\includegraphics[width=0.32\textwidth]{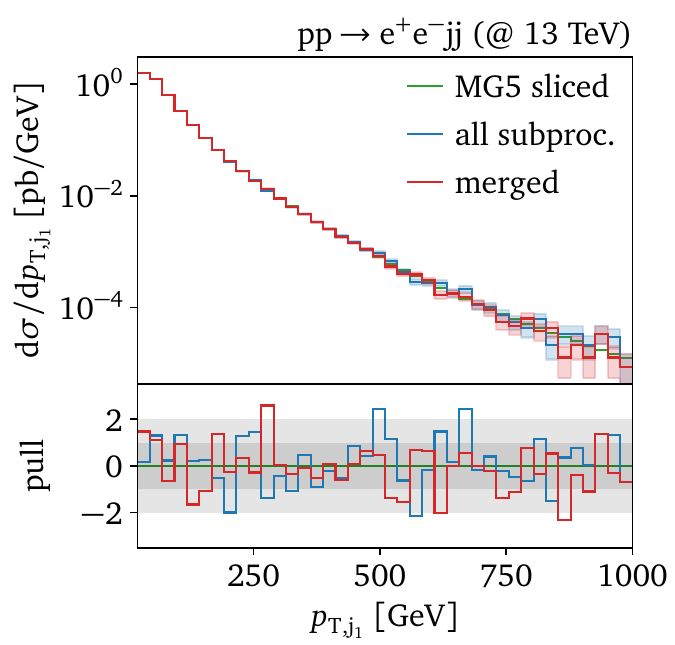}
\includegraphics[width=0.32\textwidth]{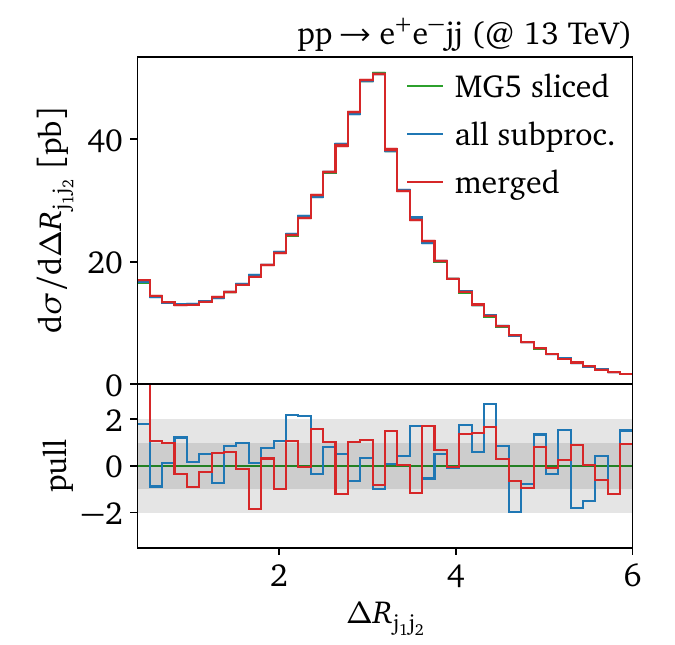}
\caption{Differential distributions for $\Pp\Pp \to \Pt\Ptbar + 2\Pj$ (top) and $\Pp\Pp \to \Pep\Pem + 2\Pj$ (bottom). Shown are the partonic center-of-mass energy or the lepton-pair invariant mass (left), the transverse momentum of the leading jet (middle), and the angular separation of the two leading jets (right). We compare unweighted \madnis events from separate and merged trainings with a sliced high-statistics \mgfive run. The lower panels show the pull of \cref{eq:pull}.}
\label{fig:validation}
\end{figure}

As observables, we choose the partonic center-of-mass energy $\sqrt{\hat s}$ for top-pair production and the invariant mass $m_{\Pep\Pem}$ of the lepton pair for Drell-Yan, as well as the transverse momentum $p_{\text{T},\mathrm{j}_1}$ of the leading jet and the angular separation $\Delta R_{\mathrm{j}_1 \mathrm{j}_2}$ between the two leading jets for both processes.
To compare the distributions bin by bin, we compute the pull
\begin{equation}
    t = \frac{\mu_{\text{\madnis}} - \mu_{\text{\mgfive}}}{\sqrt{\delta_{\text{\madnis}}^2 + \delta_{\text{\mgfive}}^2}} \;,
    \label{eq:pull}
\end{equation}
where $\mu$ denotes the cross section in a given bin and $\delta$ its statistical uncertainty.
For statistically compatible distributions, the pulls follow a standard normal distribution, so about $68\%$ and $95\%$ of the bins should lie within $|t|<1$ and $|t|<2$, respectively.
\Cref{fig:validation} shows that the distributions from both the separate and the merged trainings meet this criterion, in the bulk as well as in the tails.

\subsection{Training time and amortization}
\label{sec:training-time}

Like a \vegas grid, a trained \madnis sampler is specific to one setup, and any change of process, cuts, or model parameters requires a new training.
Because neural importance sampling needs much longer training than \vegas, see \cref{sec:merging}, its gain in unweighting efficiency only pays off once it amortizes this training time.
Here, we measure how the event throughput scales with the training time and the final-state multiplicity, and determine the number of events beyond which \madnis becomes faster than the \vegas baseline.

\subsubsection*{Setup}

We study $\Pp\Pp \to \Pt\Ptbar + n\Pj$ with $n=1,2,3$ from \cref{eq:processes}, which consist of 4, 8, and 10 subprocesses, respectively.
We either train a separate sampler for each subprocess or merge all subprocesses into a single training, as described in \cref{sec:merging}.
Apart from the training length and the merging strategy, we use the default settings of \mgseven from \cref{sec:defaults}, which fix the network size, the batch size, the buffered fraction, and the remaining hyperparameters during the \vegas pre-training and survey.
We vary the training length between 100 and $4.7 \times 10^5$ iterations on the GPU and between 100 and $4.7 \times 10^4$ iterations on the CPU, and repeat every run five times.

We run on an Nvidia A100 GPU and on 8 cores of an AMD EPYC 7452 CPU, which matches the core count of a typical LHC grid job. 
On the CPU, the matrix elements use AVX2 vectorization.
Training and event generation run on the same hardware, even though our framework also supports mixed setups such as training on a GPU and generating events on a CPU.
As figure of merit, we use the unweighted-event throughput $\dot{N}_{\text{\madnis}}$, \ie the number of unweighted events per second of generation while allowing for $0.1\%$ of the cross-section to come from over-weight events.
We compare it to the throughput $\dot{N}_{\text{\vegas}}$ of a \vegas run with \madspace on the same hardware and with otherwise identical settings.
In the limit of a large number of generated events, $N\to\infty$, the training time becomes negligible and the speed-up reduces to
\begin{equation}
\kappa_\infty=\frac{\dot{N}_{\text{\madnis}}}{\dot{N}_{\text{\vegas}}}\;,
\label{eq:speedup}
\end{equation}
which we generalize to finite $N$ in \cref{eq:speedup-total}. For both methods, we neglect the time of the common \vegas survey, which is small compared to all other contributions.

\begin{figure}[t]
\centering
\includegraphics[width=0.32\textwidth]{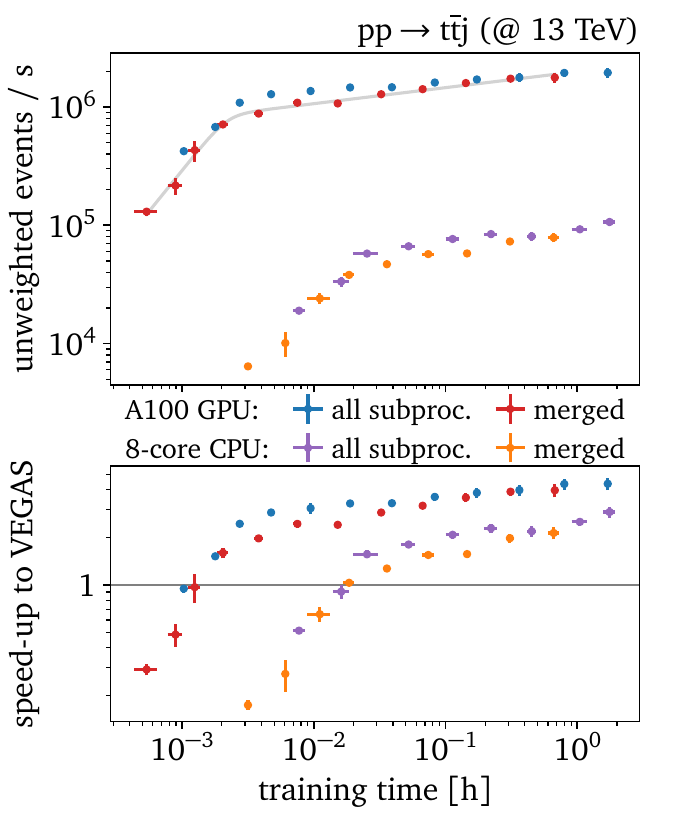}
\includegraphics[width=0.32\textwidth]{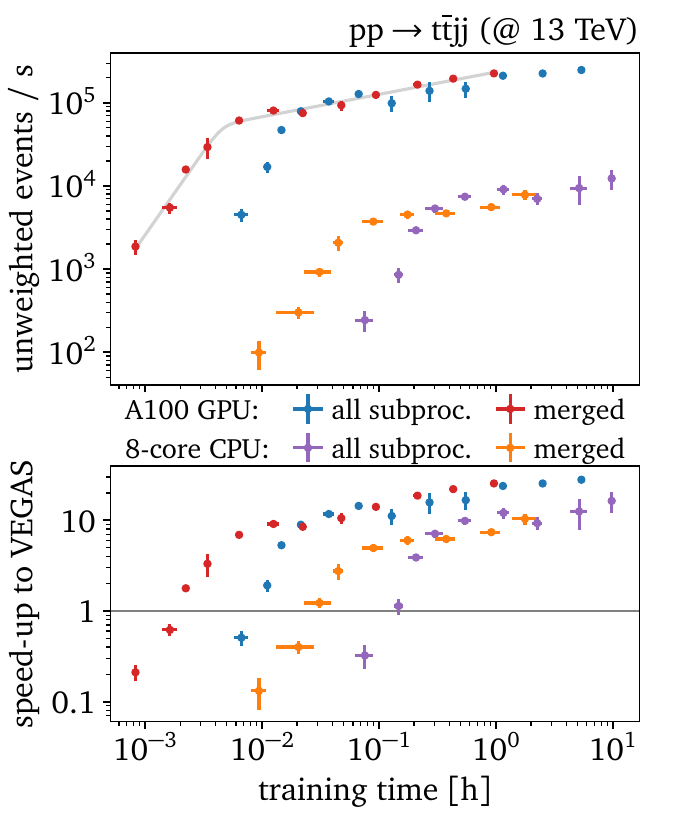}
\includegraphics[width=0.32\textwidth]{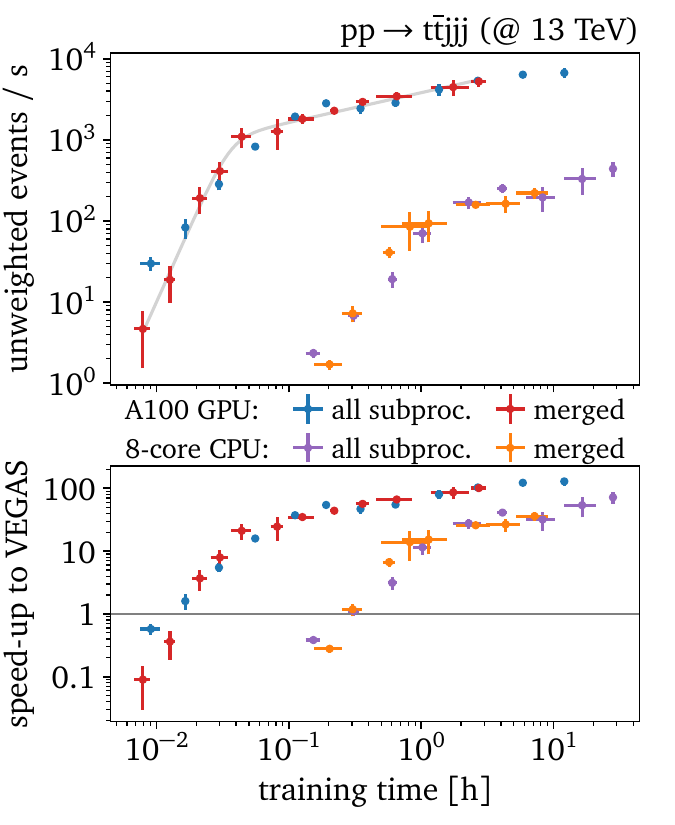}
\caption{Unweighted-event throughput (upper) and speed-up relative to \vegas on the same hardware (lower) as a function of the training time for $\Pp\Pp \to \Pt\Ptbar + n\Pj$ with $n=1,2,3$ (left to right). We show separate trainings for all subprocesses and a single merged training, each on an A100 GPU and on 8 CPU cores. Points and vertical error bars give the mean and standard deviation over five independent runs, horizontal error bars the spread in wall-clock time. The gray line shows the fit of \cref{eq:scaling-fit} to the merged GPU runs.}
\label{fig:scaling}
\end{figure}

\subsubsection*{Scaling with training time}

\Cref{fig:scaling} shows the throughput and the speed-up as a function of the training time.
For the separate trainings, every subprocess is trained with the full number of iterations, and the training time is the sum over all subprocesses.
On the GPU, the throughput first rises steeply and then enters a regime in which it grows only slowly, but does not saturate within the probed range.
We describe this behavior by two linear functions in the log-log plane, which are smoothly connected,
\begin{align}
  \log \dot{N}_{\text{\madnis}}(\ttrain) = a + b \log \ttrain - c \log\Bigl[ 1 + \exp\bigl( -d\,(\log \ttrain - \log t_0) \bigr) \Bigr] \; ,
  \label{eq:scaling-fit}
\end{align}
and fit it to the merged GPU runs.
For $\ttrain \gg t_0$ the throughput scales as $\ttrain^{b}$, and for $\ttrain \ll t_0$ as $\ttrain^{b+cd}$.
The transition happens after $t_0 \approx 8~\text{s}$, $17~\text{s}$, and $2~\text{min}$ of training for $n=1,2,3$, and beyond it the throughput grows roughly as $\ttrain^{0.1\ldots0.4}$.
\madnis reaches the \vegas throughput after $5~\text{s}$, $7~\text{s}$, and $1~\text{min}$ of GPU training for $n=1,2,3$.
For the longest trainings, the speed-up reaches $4.4$, $28$, and $130$. 
The speed-up grows with the final-state multiplicity, because the integrand on the unit hypercube becomes increasingly correlated, while the factorized \vegas grid cannot capture correlations between the integration variables, see \cref{sec:recap}.
The CPU runs follow the same pattern, but each iteration takes longer, which shifts them to longer training times by more than an order of magnitude.
To keep the wall-clock time manageable, we train them for at most a tenth of the GPU iterations, so their longest trainings reach smaller speed-ups of $3$, $16$, and $70$.
For the same number of iterations, CPU and GPU trainings reach comparable speed-ups.

For similar total training times, separate and merged trainings perform comparably.
For short trainings, merging can help substantially. This is most visible for $n=2$, where the merged training reaches a speed-up of $9$ after $40~\text{s}$ on the GPU, while the separate trainings stay at $2$.
For $n=1$, the separate trainings are slightly ahead at intermediate training times.
When to use which mode beyond $\Pt\Ptbar$ production requires further study, in line with discussion in \cref{sec:merging}.

\begin{figure}[t]
\centering
\includegraphics[width=0.2979\textwidth]{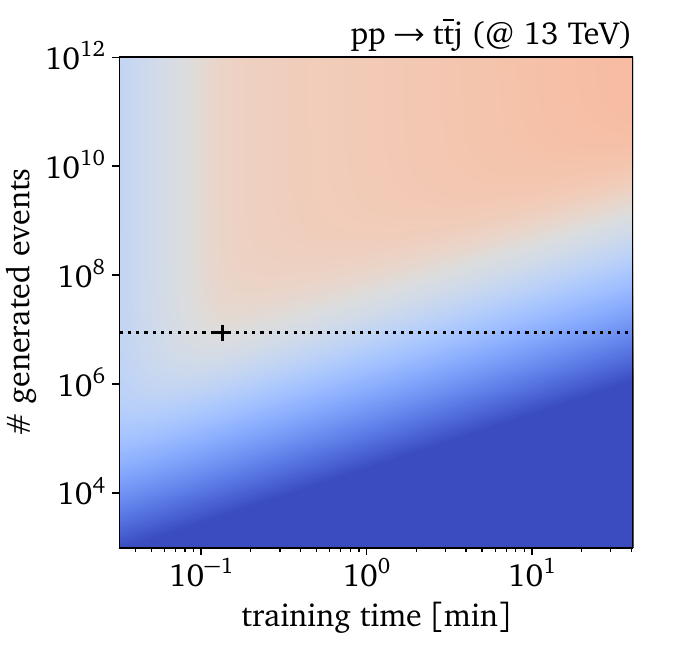}
\includegraphics[width=0.2979\textwidth]{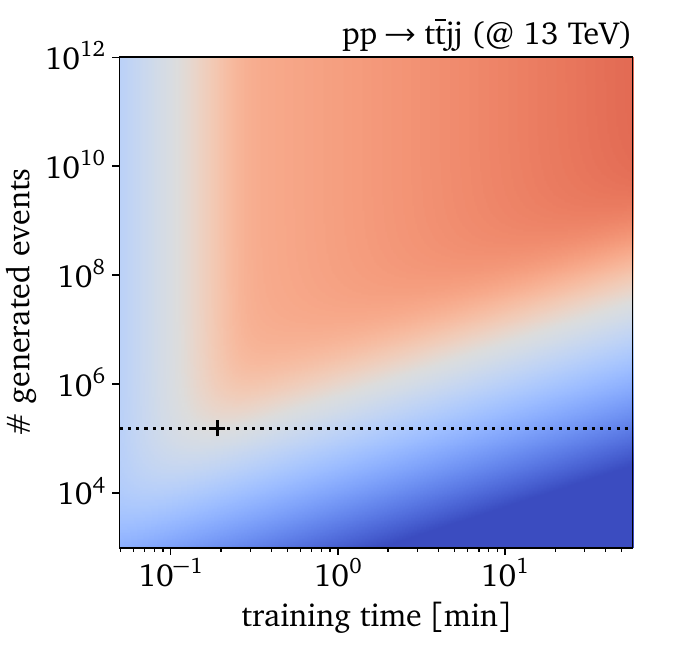}
\includegraphics[width=0.3641\textwidth]{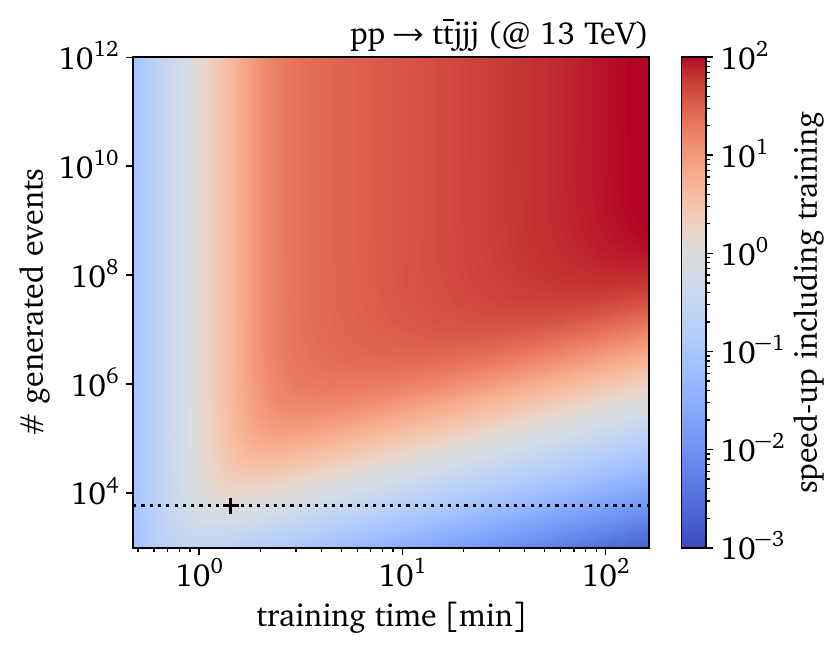}
\caption{Speed-up including the training time, \cref{eq:speedup-total}, as a function of the training time and the number of generated events for $\Pp\Pp \to \Pt\Ptbar + n\Pj$ with $n=1,2,3$ (left to right). It is derived from the fit of \cref{eq:scaling-fit} to the merged GPU runs of \cref{fig:scaling}. The cross marks the minimal number of events for which \madnis breaks even with \vegas, and the dotted line indicates this number of events.}
\label{fig:break_even}
\end{figure}

\subsubsection*{Training time amortization}

From the fit, we compute the speed-up including the training time.
Generating $N$ unweighted events takes
\begin{align}
  T_{\text{\madnis}}(N,\ttrain) = \ttrain + \frac{N}{\dot{N}_{\text{\madnis}}(\ttrain)}
  \qquad \mand \qquad
  T_{\text{\vegas}}(N) = \frac{N}{\dot{N}_{\text{\vegas}}} \; ,
  \label{eq:gen-time}
\end{align}
so the speed-up including the training time is
\begin{align}
  \kappa(N,\ttrain) = \frac{T_{\text{\vegas}}(N)}{T_{\text{\madnis}}(N,\ttrain)} \; .
  \label{eq:speedup-total}
\end{align}
For large $N$, $\kappa$ approaches $\kappa_\infty$ of \cref{eq:speedup}. For small $N$ and long trainings, the training dominates and $\kappa \ll 1$.
\madnis breaks even at $\kappa=1$, \ie starts gaining for
\begin{align}
  N_\text{gain}(\ttrain) = \frac{\ttrain\, \dot{N}_{\text{\madnis}}(\ttrain)\, \dot{N}_{\text{\vegas}}}{\dot{N}_{\text{\madnis}}(\ttrain) - \dot{N}_{\text{\vegas}}} \qquad \mand \qquad \dot{N}_{\text{\madnis}}(\ttrain) > \dot{N}_{\text{\vegas}}\; .
  \label{eq:break-even}
\end{align}
\Cref{fig:break_even} shows $\kappa(N,\ttrain)$, and the cross marks the minimum of $N_\text{gain}(\ttrain)$.
\madnis breaks even at $9 \times 10^6$, $1.5 \times 10^5$, and $6 \times 10^3$ events for $n=1,2,3$, after $8~\text{s}$, $11~\text{s}$, and $1.4~\text{min}$ of training.
The higher the multiplicity, the earlier \madnis pays off, and for $\Pt\Ptbar + 3\Pj$ a few thousand events suffice.
For $n=1$, \vegas is already efficient, so only large samples benefit from using \madnis.
Beyond the break-even point, the optimal training time grows with $N$. The training length therefore has to follow the requested number of events, which motivates the defaults of \cref{sec:defaults}.

\Cref{tab:scaling} summarizes three working points of the merged GPU and CPU trainings together with the \vegas baseline.
Here, $\epsilon_\text{uw}$ denotes the unweighting efficiency.
Already after $10^3$ iterations, which take between 17~s and 200~s on the GPU to train, \madnis generates unweighted events faster than \vegas by factors of 2.9, 8.9, and 16 for $n=1,2,3$.
After $4.7 \times 10^5$ iterations, the unweighting efficiency improves by factors of 6.7, 25, and 100, and the throughput by factors of 4.2, 28, and 130.

\begin{table}[t]
\centering
\renewcommand{\arraystretch}{1.2}
\begin{small}
\setlength{\tabcolsep}{4pt}
\begin{tabular}{l|rcS[table-format=1.3]|rcS[table-format=1.3]|rcS[table-format=1.3]}
\toprule
 & \multicolumn{3}{c|}{$\mathrm{p}\mathrm{p}\to\mathrm{t}\overline{\mathrm{t}}+1\,\mathrm{j}$} & \multicolumn{3}{c|}{$\mathrm{p}\mathrm{p}\to\mathrm{t}\overline{\mathrm{t}}+2\,\mathrm{j}$} & \multicolumn{3}{c}{$\mathrm{p}\mathrm{p}\to\mathrm{t}\overline{\mathrm{t}}+3\,\mathrm{j}$} \\
 & $t_\mathrm{train}$ & unw./s & {$\epsilon_\mathrm{uw}$} & $t_\mathrm{train}$ & unw./s & {$\epsilon_\mathrm{uw}$} & $t_\mathrm{train}$ & unw./s & {$\epsilon_\mathrm{uw}$} \\
\midrule
GPU, 1k it. & 17\,s & $1.3 \times 10^{6}$ & 0.38 & 78\,s & $7.9 \times 10^{4}$ & 0.14 & 200\,s & $8.3 \times 10^{2}$ & 0.021 \\
CPU, 1k it. & 190\,s & $6.6 \times 10^{4}$ & 0.42 & 18\,min & $5.3 \times 10^{3}$ & 0.14 & 1.0\,h & $7.0 \times 10^{1}$ & 0.019 \\
GPU, 47k it. & 10\,min & $1.7 \times 10^{6}$ & 0.54 & 33\,min & $1.5 \times 10^{5}$ & 0.25 & 1.4\,h & $4.2 \times 10^{3}$ & 0.089 \\
CPU, 47k it. & 1.8\,h & $1.1 \times 10^{5}$ & 0.64 & 9.8\,h & $1.2 \times 10^{4}$ & 0.33 & 28\,h & $4.4 \times 10^{2}$ & 0.096 \\
GPU, 470k it. & 1.7\,h & $1.9 \times 10^{6}$ & 0.64 & 5.4\,h & $2.5 \times 10^{5}$ & 0.38 & 12\,h & $6.7 \times 10^{3}$ & 0.160 \\
\midrule
GPU, VEGAS & -- & $4.5 \times 10^{5}$ & 0.095 & -- & $8.9 \times 10^{3}$ & 0.015 & -- & $5.2 \times 10^{1}$ & 0.0016 \\
CPU, VEGAS & -- & $3.7 \times 10^{4}$ & 0.10 & -- & $7.5 \times 10^{2}$ & 0.013 & -- & $6.1 \times 10^{0}$ & 0.0010 \\
\bottomrule
\end{tabular}
\end{small}
\caption{Training time $\ttrain$, unweighted-event throughput $\dot N$, and unweighting efficiency $\epsilon_\text{uw}$ for $\Pp\Pp \to \Pt\Ptbar + n\Pj$ with $n=1,2,3$, after $10^3$, $4.7 \times 10^4$, and $4.7 \times 10^5$ training iterations on an A100 GPU and on 8 CPU cores. The last two rows show the \vegas baseline on the same hardware. The unweighting efficiency allows for $0.1\%$ of the cross section to come from overweight events.}
\label{tab:scaling}
\end{table}

\subsection{The process zoo}
\label{sec:process-zoo}

Finally, we run \madnis with its default settings on a broad set of LHC processes,
\begin{alignat}{2}
  \text{multi-jet:} \qqquad
    & \Pp\Pp \to 3\Pj \;, \qquad
    && \Pp\Pp \to 4\Pj \;. \notag \\
  \text{Drell-Yan and jets:} \qqquad
    & \Pp\Pp \to \Pl^+\Pl^- + 2\Pj \;, \qquad
    && \Pp\Pp \to \Pl^+\Pl^- + 3\Pj \;. \notag \\
  \text{$\PW$ and jets:} \qqquad
    & \Pp\Pp \to \PWpm + 2\Pj \;,
    && \Pp\Pp \to \PWpm + 3\Pj \;. \notag \\
  \text{top quarks:} \qqquad
    & \Pp\Pp \to \Pt\Ptbar + 2\Pj \;,
    && \Pp\Pp \to \Pt\Ptbar + 3\Pj \;, \notag \\
    & \Pp\Pp \to \Pt\Ptbar\Pt\Ptbar \;, && \Pp\Pp \to \Pt\Ptbar \to \Pb\bar{\Pb}\,\Pl^+\Pl^-\nu\bar{\nu} \;. \notag \\
  \text{Higgs (VBF):} \qqquad
    & \Pp\Pp \to \PH\Pj\Pj \;,
    && \Pp\Pp \to \PH\PH\Pj\Pj \;. \notag \\
  \text{electroweak:} \qqquad
    & \Pp\Pp \to \PWp\PWm\Pj\Pj \;,
    && \Pp\Pp \to \PWp\PWm\PWp \;.
  \label{eq:zoo}
\end{alignat}
For $\PWp\PWm\Pj\Pj$, $\PH\Pj\Pj$, and $\PH\PH\Pj\Pj$, we only include purely electroweak diagrams, so these processes describe vector-boson scattering and vector-boson fusion, respectively.
All particles are generated on-shell without decays unless written otherwise.
For every process, we include all partonic subprocesses, use the setup of \cref{eq:cuts}, and generate $10^6$ unweighted events.
For the high-multiplicity processes $\Pp\Pp \to \Pt\Ptbar + 3\Pj$ and $\Pp\Pp \to \PWpm + 3\Pj$, we generate $10^5$ unweighted events to limit the computing time, which does not affect the speed-up $\kappa_\infty$ since it excludes the training.
We train \madnis separately for each subprocess with $2 \times 10^5$ iterations and otherwise use the defaults of \cref{sec:defaults}.
This is much longer than necessary for the simpler processes, but it shows the speed-ups that are possible after a long training.
We run both \madnis and \vegas on the same Nvidia H100 GPU, so the speed-up $\kappa_\infty$ of \cref{eq:speedup} measures the gain from the sampling method alone and not from the hardware.
For each process, we repeat the comparison for two or three pairs of random seeds.

\begin{figure}[t]
\centering
\includegraphics[width=\linewidth]{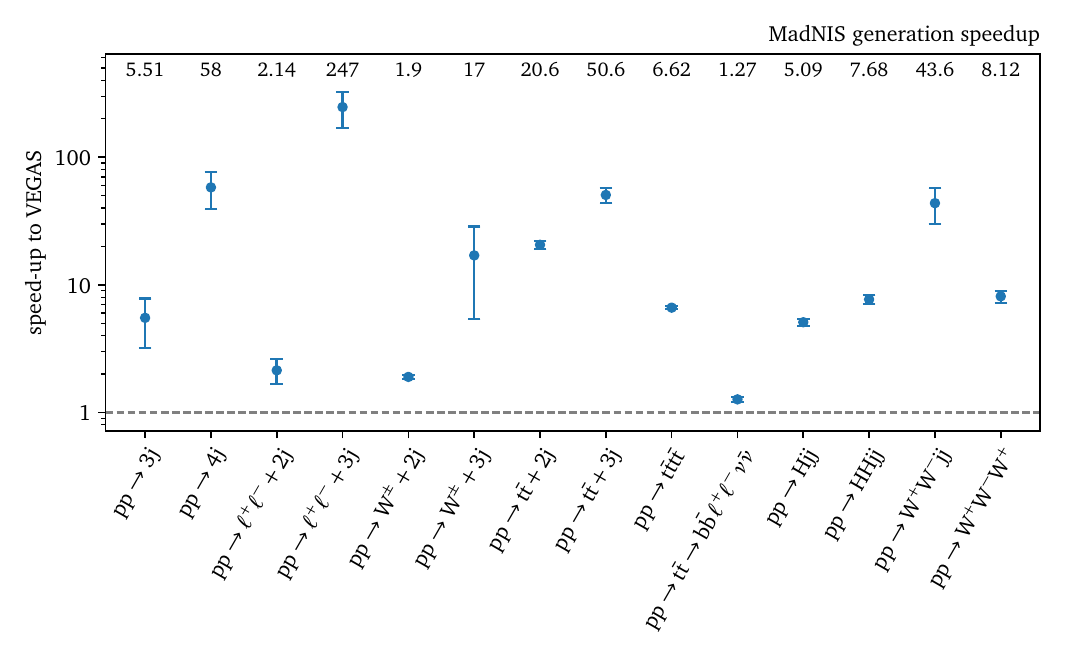}
\caption{Speed-up $\kappa_\infty$ of \madnis relative to \vegas, \cref{eq:speedup}, for the processes of \cref{eq:zoo}, with training and survey times excluded. Both methods run on the same Nvidia H100 GPU. Each dot corresponds to one of three pairs of random seeds. The boxes show the interquartile range, the horizontal lines and the numbers above the boxes the median, and the whiskers the full range.}
\label{fig:madnis_vs_vegas}
\end{figure}

\Cref{fig:madnis_vs_vegas} shows the resulting speed-ups.
In the median, \madnis is faster than \vegas for every process, with speed-ups between 1.27 and 247.
The gain grows strongly with the final-state multiplicity.
One additional jet increases it from 5.5 to 58 for multi-jet production, from 20 to 50 for $\Pt\Ptbar$ plus jets, from 1.9 to 17 for $\PWpm$ plus jets, and from 2.1 to 247 for $\Pl^+\Pl^-$ plus jets.
For high multiplicities, the integrand becomes increasingly correlated, and the number of integration channels grows. \madnis captures the correlations and drops channels that do not contribute, which makes the remaining training and sampling more efficient.
Electroweak vector-boson scattering, $\PWp\PWm\Pj\Pj$, gains a factor 44.
The smallest gains appear for processes in which the analytic phase-space mappings already describe the integrand well.
For top-pair production with leptonic decays in the narrow-width approximation, the Breit-Wigner mappings capture the resonance structure, and \vegas leaves little to improve, so \madnis gains only a factor 1.27.
The variation between seeds is small for most processes.

\clearpage
\section{Outlook}
\label{sec:outlook}

We have made neural importance sampling with \madnis a production-ready default for leading-order event generation in \mg, applicable to generic proton-proton processes with all their partonic subprocesses.
Importance sampling of symmetry permutations and flavor configurations with a PDF prior covers the discrete degrees of freedom of the integrand, and merging matrix elements, channels, and subprocesses reduces the number of independent trainings.
Buffered training on partially unweighted events stabilizes the training, concurrent training and generation keep the hardware busy, and automated defaults derived from a short \vegas survey remove the need for extensive process-specific tuning. All developments are publicly available in the alpha release of \mgseven~\cite{MG7:v0.2.1}, which will succeed \mgfive as the production version of \mg.

With these defaults, \madnis is faster than \vegas by factors between 1.27 and 247 across a broad range of LHC processes, with the largest gains at high final-state multiplicities.
The gain grows with the multiplicity, and for $\Pt\Ptbar + 3\Pj$ the training cost already amortizes after 6000 unweighted events.
At the same time, the generated distributions agree with an independent high-statistics baseline, in the bulk as well as in the tails.
Learning the flavor structure on top of the PDF prior gives an additional gain whenever the matrix element depends on the flavor configuration.

These gains are largest for intermediate final-state multiplicities, which is the regime targeted in this paper.
For low multiplicities, \vegas combined with the analytic mappings of \madspace is already efficient, and the hardware acceleration of phase-space generation and matrix elements in \madspace and \madmatrix~\cite{Heimel:2026hgp,Valassi:2023yud} becomes the dominant source of speed-up, which we leave to the full \mgseven release.
For very high multiplicities, the evaluation of full-color matrix elements becomes the bottleneck, which leading-color event generation with \amplicol~\cite{Frederix:2024uvy,Frederix:2026ejl} addresses in a complementary way.

For the LHC, faster event generation can be spent in two ways.
The same samples can be generated with a smaller computing footprint, which addresses the growing gap between the required simulation and the available budget for the HL-LHC.
Alternatively, the same budget yields larger samples, in particular for high-multiplicity final states, which enter many background predictions at LO through multi-jet merging and where limited Monte Carlo statistics in the tails often restrict the sensitivity of an analysis.
Because no tuning is required, this gain is available to every \mg user, from large production campaigns to smaller phenomenological studies, whenever the requested number of events is large enough to amortize the training.

\clearpage
\subsection*{Acknowledgments}

We thank the full \mg team for their continuous support and valuable input throughout this work.
TH is supported by the PDR-Weave grant FNRS-DFG numéro T019324F (40020485).
OM, TH and the madgraph collaboration are supported by FRS-FNRS (Belgian National Scientific Research Fund) IISN
projects 4.4503.16 (MaxLHC). 
The work done by DM is supported by the Eric \& Wendy Schmidt Fund for Strategic Innovation through the CERN Next Generation Triggers project (grant agreement SIF-2023-004).
Computational resources have been provided by the Consortium des Équipements de Calcul Intensif (CÉCI), funded by the Fonds de la Recherche Scientifique de Belgique (F.R.S.-FNRS) under Grant No. 2.5020.11 and by the Walloon Region.
We also acknowledge the usage of computing resources hosted at CERN through the Next Generation Triggers project (grant agreement SIF-2023-004).
This article/publication is based upon work from COST Actions MLQC4FC (CA24146) supported by
COST (European Cooperation in Science and Technology). This project is partly funded by HADRON-2030 which receives funding from the European Union’s Horizon Europe research and innovation programme under Grant Agreement No. 101292522.

\subsection*{Disclaimer}
\begin{figure}[ht] \centering \begin{minipage}{0.35\textwidth} \includegraphics[width=\linewidth]{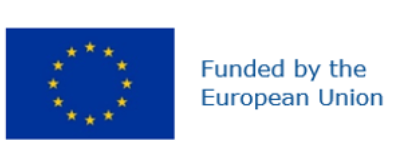} \end{minipage} \hfill \begin{minipage}{0.6\textwidth} Funded by the European Union. Views and opinions expressed are however those of the author(s) only and do not necessarily reflect those of the European Union or the European Research Executive Agency (REA). Neither the European Union nor the granting authority can be held responsible for them. \end{minipage} \end{figure}

\clearpage
\appendix
\section{Hyperparameters}

\begin{table}[h!]
    \setlength{\tabcolsep}{10pt}
    \renewcommand{\arraystretch}{1.5}
    \newcommand{\stackcell}[1]{{\renewcommand{\arraystretch}{1}\begin{tabular}[c]{@{}l@{}}#1\end{tabular}}}
    \centering
    \begin{small} \begin{tabular}{l|c}
        \toprule
        Hyperparameter & value\\
        \midrule
        \multicolumn{2}{c}{\textit{process-dependent}} \\
        \midrule
        \stackcell{Hidden nodes (flow, discrete flow,\\ channel weights)}
            & $\min\left\{\max\left\{32 \left\lfloor \frac{7\,\text{RSD}}{32} \right\rfloor + 64,\; 64\right\},\; 256\right\}$ \\[4pt]
        Layers per flow subnet
            & 3 if $\text{RSD} < 32$, 4 otherwise \\
        Batch size per channel
            & $\min\left\{\max\left\{32 \left\lfloor \frac{7\,\text{RSD}}{32} \right\rfloor + 64,\; 128\right\},\; 512\right\}$ \\[4pt]
        Learning rate
            & $\min\left\{\max\left\{3\cdot10^{-4} + 7\cdot10^{-4}\,\frac{10^4 - B}{8000},\; 3\cdot10^{-4}\right\},\; 10^{-3}\right\}$ \\[4pt]
        \stackcell{Fraction of iterations with\\ frozen channel weights}
            & $\max\left\{0.33,\; 1 - \frac{10^4}{B}\right\}$ \\[4pt]
        Coupling blocks
            & $\lceil \log_2 d \rceil$ \\
        Channel dropping threshold
            & $10^{-2} \cdot \min\left\{\frac{1000}{b},\; 1\right\}$ \\[4pt]
        \midrule
        \multicolumn{2}{c}{\textit{fixed}} \\
        \midrule
        Loss & stratified variance \\
        Activation function & leaky ReLU \\
        RQ spline bins & 10 \\
        \stackcell{Layers (discrete flow,\\ channel weight network)} & 3 \\
        Constant batch size offset & 512 \\
        \stackcell{Fraction of points sampled\\ uniformly across channels} & 0.5 \\
        Optimizer & Adam \\
        Adam $(\beta_1, \beta_2, \epsilon)$ & $(0.9,\, 0.999,\, 10^{-8})$ \\
        Weight decay & $10^{-4}$ \\
        LR schedule & cosine annealing \\
        Gradient clipping threshold & $3\cdot 10^{-3} \times \text{number of parameters}$ \\
        \stackcell{Softclip threshold\\ for variance loss} & 30 \\
        Buffer capacity per channel & 60000 \\
        \stackcell{Minimum buffer size before\\ buffered training begins} & 10000 \\
        Fraction of buffered iterations & 0.8 \\
        \stackcell{Initial iterations without\\ saving samples to buffer} & 1000 \\
        Buffer unweighting quantile & 0.95 \\
        \bottomrule
    \end{tabular} \end{small}
    \caption{Fixed and process-dependent \madnis hyperparameters. RSD denotes the relative standard deviation determined in the \vegas survey. $B$ denotes the number of training iterations, $b$ the current iteration, and $d$ the dimension of the phase space.}
    \label{tab:hyperparameters}
\end{table}

\section{Per-process generation performance}
\label{app:madnis-generation-performance}

\cref{tab:madnis-generation-performance} shows the generation performance for each selected process. 
The speed-up is the mean VEGAS generation wall time divided by the mean MadNIS generation wall time, with the two methods treated as independent samples.
The throughput is the number of target unweighted events divided by the generation wall time, and the unweighting efficiency is calculated as the target count divided by the unweighted event count after cuts.
We report the arithmetic mean of each per-run MadNIS quantity.
The timing comparison covers event generation only, with the training time shown separately.

\begin{table}[h!]
  \centering
  \small
  \setlength{\tabcolsep}{4pt}
  \begin{tabular}{lrrrr}
    \toprule
    Process & \shortstack{Speedup} & \shortstack{Throughput [unw./s]} & \shortstack{$\epsilon_{\mathrm{uw}}$} & \shortstack{$t_\textup{train}$} \\
    \midrule
    $\Pp\Pp \to 3\Pj$ & 5.51 & $3.82\times10^{5}$ & 0.09 & 1.5 h \\
    $\Pp\Pp \to 4\Pj$ & 58 & $4.53\times10^{3}$ & 0.00524 & 4.4 h \\
    $\Pp\Pp \to \Pl^+\Pl^- + 2\Pj$ & 2.14 & $4.60\times10^{4}$ & 0.0127 & 2.9 h \\
    $\Pp\Pp \to \Pl^+\Pl^- + 3\Pj$ & 247 & $1.69\times10^{4}$ & 0.0136 & 6.7 h \\
    $\Pp\Pp \to \PWpm + 2\Pj$ & 1.9 & $4.08\times10^{5}$ & 0.0716 & 3.4 h \\
    $\Pp\Pp \to \PWpm + 3\Pj$ & 17 & $9.60\times10^{4}$ & 0.0524 & 8.8 h \\
    $\Pp\Pp \to \Pt\Ptbar + 2\Pj$ & 20.6 & $3.70\times10^{5}$ & 0.291 & 2.3 h \\
    $\Pp\Pp \to \Pt\Ptbar + 3\Pj$ & 50.6 & $7.08\times10^{3}$ & 0.0462 & 4.1 h \\
    $\Pp\Pp \to \Pt\Ptbar\Pt\Ptbar$ & 6.62 & $9.32\times10^{5}$ & 0.533 & 22 min \\
    $\Pp\Pp \to \Pt\Ptbar \to \Pb\bar{\Pb}\,\Pl^+\Pl^-\nu\bar{\nu}$ & 1.27 & $2.04\times10^{6}$ & 0.415 & 14 min \\
    $\Pp\Pp \to \PH\Pj\Pj$ & 5.09 & $3.72\times10^{6}$ & 0.565 & 50 min \\
    $\Pp\Pp \to \PH\PH\Pj\Pj$ & 7.68 & $1.01\times10^{6}$ & 0.241 & 86 min \\
    $\Pp\Pp \to \PWp\PWm\Pj\Pj$ & 43.6 & $5.74\times10^{4}$ & 0.0645 & 3.7 h \\
    $\Pp\Pp \to \PWp\PWm\PWp$ & 8.12 & $3.38\times10^{6}$ & 0.7 & 6.0 min \\
    \bottomrule
  \end{tabular}
  \caption{MadNIS performance for each process.
  Speed-up is the ratio in between the geometric means of VEGAS and MadNIS generation times using all completed runs for each method independently.
  The remaining columns are arithmetic means over completed MadNIS runs: generation throughput $N_{\textup{target}}/t_{\textup{gen}}$, unweighting efficiency $\epsilon_{\textup{uw}}=N_{\textup{target}}/N_{\textup{after cuts}}$, and training wall time.
  Throughput excludes training and survey time.
  Runs use NVIDIA H100 GPU and MadGraph7 v0.2.1.}
  \label{tab:madnis-generation-performance}
\end{table}
\clearpage
\bibliography{refs}
\end{document}